\documentclass[aps,prd,reprint,preprintnumbers,superscriptaddress,nofootinbib,floatfix]{revtex4-1}

\usepackage[T1]{fontenc}
\usepackage[utf8]{inputenc}
\usepackage{microtype}

\usepackage{amsmath,amssymb,amsfonts,bm,slashed,dcolumn}
\numberwithin{equation}{section}

\usepackage{graphicx}
\usepackage{tabularx}
\usepackage{subfigure}  
\usepackage{placeins}

\usepackage{xcolor}
\usepackage{url}
\usepackage{comment}

\usepackage[
  colorlinks=true,
  linkcolor=blue,
  citecolor=blue,
  urlcolor=blue
]{hyperref}

\newcommand{\be}{\begin{equation}}
\newcommand{\ee}{\end{equation}}
\newcommand{\ba}{\begin{eqnarray}}
\newcommand{\ea}{\end{eqnarray}}

\begin{document}
\preprint{JCAP \textbf{10} (2026) 043, DOI: 10.1088/1475-7516/2026/10/043}

\title{Energy extraction from NED-deformed rotating black holes via the Comisso--Asenjo reconnection process}

\author{Marco Figliolia}
\email{mfigliolia@unisa.it}
\affiliation{Dipartimento di Fisica ``E.R. Caianiello'', Universit\`{a} degli Studi di Salerno, Via Giovanni Paolo II, 132, 84084 Fisciano (SA), Italy}

\author{Gaetano Lambiase}
\email{glambiase@unisa.it}
\affiliation{Dipartimento di Fisica ``E.R. Caianiello'', Universit\`{a} degli Studi di Salerno, Via Giovanni Paolo II, 132, 84084 Fisciano (SA), Italy}
\affiliation{INFN Sezione di Napoli, Gruppo Collegato di Salerno, Via Giovanni Paolo II 132, I-84084 Fisciano (SA), Italy}

\author{A.~\"Ovg\"un}
\email{aliovgun@gmail.com}
\affiliation{Physics Department, Eastern Mediterranean University, Famagusta, 99628 North Cyprus, via Mersin 10, T\"urkiye}

\author{R.~C.~Pantig}
\email{reggie.pantig04@gmail.com}
\affiliation{Physics Department, School of Foundational Studies and Education, Map\'ua University, 658 Muralla St., Intramuros, Manila 1002, Philippines}

\begin{abstract}
We study rotating black holes in general relativity coupled to nonlinear electrodynamics (NED), focusing on the axisymmetric solution of \cite{GhoshWalia2021} with deformation parameter \(g\).
On the spherical seed, weak--field lensing via the Gauss--Bonnet method and the shadow radius yield a spin-insensitive bound by enforcing a conservative \(\sim10\%\) tolerance on the Sgr~A* ring size, namely \(g/M\!\lesssim\!1.26\). In the eikonal regime we derive analytic quasinormal-mode shifts (even in \(g\)) and obtain an independent ceiling consistent with the shadow constraint \cite{Churilova:2019jqx}.
For the rotating geometry, we provide closed-form ZAMO scalars, chart horizons and ergoregion, and analyze equatorial geodesics (photon orbits and ISCO). We then formulate in the ZAMO frame the Comisso--Asenjo reconnection channel \cite{ComissoAsenjoPRD2021}, identify the negative-energy window, and integrate the extracted power over the allowed radii; from the tolerated fractional departure from the Kerr power we define a spin-dependent extraction bound \(g_\delta(a\mid\sigma_0,\xi)\).
Taken together, the QNM/shadow ceiling and the extraction bound appreciably narrow the admissible region for \(g/M\) in the \((a/M,\,g/M)\), so even within our deliberately simplified, single-layer equatorial setup, the two complementary probes already provide informative constraints on NED deformations, testable with present data and upcoming horizon-scale/ringdown campaigns.
\end{abstract}

\maketitle

\section{Introduction}\label{sec:level1}
Black holes (BHs) stand as one of the most profound and robust predictions of Einstein's general relativity (GR), offering an unparalleled arena to probe the gravitational interaction in its most extreme regime. While GR has passed all weak-field tests with remarkable precision, it is in the near-horizon environment of astrophysical BHs—characterized by strong spacetime curvature and high-energy plasma dynamics—that deviations from classical behavior might emerge. The first image of a black hole shadow by the Event Horizon Telescope (EHT) \cite{EHT2019,Akiyama2019,GRAVITY2020}, alongside the detection of gravitational waves from binary BH mergers, has provided an observational window into these strong-field regions. Yet, the theoretical foundations of black hole physics remain incomplete, with fundamental challenges such as the nature of the singularity, the quantum origin of black hole entropy, and the limits of GR at high energies still unresolved \cite{HawkingEllis,Wald2001,BardeenCarterHawking1973}.
\\

A promising avenue to address some of these issues involves modifying the classical action of GR by coupling it to nonlinear matter fields. Among such frameworks, nonlinear electrodynamics (NED) has attracted significant interest \cite{Allahyari:2019jqz,Verbin:2024ewl,FradkinTseytlin1985,Ayon-Beato:2000mjt,Ayon-Beato:1999kuh,Gibbons:1995cv,Fan:2016hvf,Novello:1999pg,Hassaine:2008pw,Maeda:2008ha,Kruglov:2017fck,Kruglov:2016ymq,Dymnikova:2004zc}. Originally introduced by Born and Infeld to eliminate the divergence in the self-energy of point charges \cite{BornInfeld1934}, NED reemerged in the context of low-energy effective actions in string theory and D-brane physics \cite{FradkinTseytlin1985,Tseytlin1999}, where it regularizes field strengths in strongly curved regimes. From a phenomenological perspective, NED couplings can naturally encode quantum corrections to electrodynamics near Planckian scales, offering a controlled deformation of the classical Maxwell action that becomes significant only at high field intensities.
When embedded in a gravitational context, NED fields modify the effective stress-energy tensor that sources the Einstein equations, thus leading to deformations of black hole spacetimes. These modifications may regularize the curvature invariants at the origin or alter the causal structure in the deep infrared, allowing for singularity resolution or nontrivial horizon topologies. Such models have led to the construction of a wide class of regular black hole solutions in GR coupled to NED sources, including the Bardeen black hole interpreted as a magnetic monopole in a specific NED model \cite{Ayon-Beato:2000mjt}. More recently, in Refs.~\cite{LambiasePRD2008,LambiaseEPJC2019} it has been explored how NED modifies the thermodynamic behaviour and quantum properties of black holes, including its effects on Hawking radiation and entropy-area relations.
\\

Of particular interest is the work of Ghosh and Walia \cite{GhoshWalia2021}, who constructed an exact rotating BH solution in GR coupled to a magnetic NED field using a refined Newman–Janis algorithm. Their solution smoothly deforms the Kerr metric, introducing a dimensionful parameter \(g\) controlling the strength of the nonlinear effects. Physically, this parameter encapsulates the scale at which nonlinear electromagnetic interactions become dominant over Maxwellian behavior, introducing curvature corrections that remain negligible at large distances but become significant near the horizon. The resulting spacetime retains asymptotic flatness and axisymmetry, but exhibits nontrivial modifications to the horizon structure, the topology of the ergoregion, and the location of characteristic surfaces such as the innermost stable circular orbit (ISCO) and the static limit. These features make it a promising candidate to explore gravitational dynamics and energetic phenomena in geometries that deviate minimally—but meaningfully—from the Kerr paradigm.
\\

These geometric features are not just of mathematical interest—they have deep implications for the dynamics of energy extraction and the astrophysical observables associated with high-energy BH environments \cite{Abdujabbarov:2011af,Oteev:2016fbp,Toshmatov:2015gna,Banados:2009pr}. Since the pioneering proposal by Penrose \cite{Penrose1969,ChristodoulouRuffini1971} that rotational energy could be extracted from a BH via particle splitting in the ergosphere, various mechanisms have been proposed and refined. The Blandford–Znajek (BZ) process \cite{BlandfordZnajek1977} models energy extraction via magnetic fields threading the event horizon, converting spin energy into outgoing Poynting fluxes. Magnetohydrodynamic (MHD) extensions \cite{Tchekhovskoy2011} include the plasma inertia and electromagnetic backreaction, yielding rich and highly efficient energy release channels. The theoretical basis for rotational energy extraction has deep historical roots, notably developed in a sequence of works by Dadhich and collaborators \cite{Dadhich1984a,Dadhich1984b,Dadhich1985a,Dadhich1986,Dadhich1989}, and later refined in more recent astrophysical contexts \cite{Dadhich2018,Dadhich2019,Dadhich2020}.
\\

More recently, Comisso and Asenjo introduced a novel mechanism in which magnetic reconnection in the ergosphere plays a central role \cite{ComissoAsenjoPRD2021}. In their model, reconnection-driven plasmoids are accelerated in opposite directions: one stream escapes to infinity with positive energy, while the other, having acquired negative energy-at-infinity, is absorbed by the black hole. This process satisfies the same energetic conditions as the Penrose process but is mediated by field-driven plasma dynamics rather than mechanical particle splitting. The efficiency of this Comisso–Asenjo (CA) process depends strongly on the geometry of the ergoregion, the orientation of the reconnection layer, and the plasma enthalpy. Importantly, it establishes a link between magnetic field topology and the extraction of gravitational energy—a connection that becomes even more relevant in non-Maxwellian contexts like NED. Recently, magnetic reconnection has also been studied in more exotic geometries, such as Kerr–Taub–NUT spacetimes, where the presence of a gravitomagnetic monopole modifies the plunging region and affects the efficiency of reconnection-based extraction \cite{Cheng2024}.
Subsequent analyses have extended the CA framework to more general backgrounds, including Kerr–Newman black holes, BHs surrounded by perfect fluid dark matter, and rotating solutions in alternative gravity theories \cite{Rodriguez2024,ZhangPRD2023,Carleo2023}.
These studies demonstrate that even mild deformations of the Kerr geometry can significantly affect the conditions for energy extraction, especially by modifying the angular velocity of the zero angular momentum observer (ZAMO), the location of the static limit, and the angular profile of the ergosphere. In these scenarios, the extracted power, the critical conditions for reconnection, and the evolution of the black hole spin can differ from the standard predictions, potentially offering observational signatures of non-Kerr metrics . Moreover, recent work has shown that nonlinearities in the plasma and metric may couple in nontrivial ways, enhancing or suppressing the reconnection efficiency depending on the local geometry and flow conditions.
\\

In this work, we analyze energy extraction in the rotating NED black–hole geometry of Ghosh and Walia \cite{GhoshWalia2021}, with the aim of quantifying how the deformation parameter \(g\) imprints on strong–field energetics. Our first step is a geometric calibration: using the spherical sector as a conservative proxy, we compute the eikonal quasinormal modes (QNMs) and obtain analytic frequency and damping shifts that scale with even powers of \(g/M\). Requiring the ringdown spectrum to remain within Kerr–consistent tolerances furnishes a uniform, microphysics–agnostic ceiling \(g_{\rm QNM}\) that we use as a prior in the rotating analysis. In parallel, we exploit the shadow size to anchor that ceiling numerically; enforcing a \(\sim 10\%\) tolerance on the ring diameter yields a conservative, spin–independent upper limit on \(g/M\) that we adopt across our parameter scans.
\\

On this calibrated backbone, we formulate the Comisso–Asenjo (CA) reconnection channel \cite{ComissoAsenjoPRD2021} in the ZAMO frame of the rotating NED metric. We derive the energy at infinity for the co– and counter–rotating outflows, identify the extraction window inside the ergoregion, and integrate the corresponding per–enthalpy energy density to obtain the total power. We then compare the NED prediction at fixed spin with the Kerr benchmark through a fractional power discrepancy and define from it a spin–resolved upper bound \(g_\delta(a\mid\sigma_0,\xi)\). The objective is methodological rather than to deliver the tightest constraint: in a controlled, transparent setting, reconnection–driven extraction can be turned into an operative upper limit on \(g/M\) that complements the uniform QNM/shadow ceiling and isolates the role of frame dragging and ergoregion geometry.


\section{\label{sec:level2} NED Black hole solution}

We work in geometrized units \(G=c=1\). The action for gravity minimally coupled to nonlinear electrodynamics (NED) is
\begin{equation}\label{actionG}
S=\int d^4x\,\sqrt{-g}\,\bigg[\;\frac{R}{16\pi}\;-\;\frac{1}{4\pi}\,\mathcal{L}_{\text{NED}}\bigg],
\end{equation}
where \(\mathcal{L}_{\text{NED}}=\mathcal{L}(\mathcal{F},\mathcal{G})\) is a scalar function of the electromagnetic invariants
\begin{equation}
\mathcal{F}\equiv \tfrac14 F_{\mu\nu}F^{\mu\nu},\qquad
\mathcal{G}\equiv \tfrac14 F_{\mu\nu}\,\tilde F^{\mu\nu}.
\end{equation}
Here \(F_{\mu\nu}=\partial_\mu A_\nu-\partial_\nu A_\mu\) and we denote by \(\tilde F^{\mu\nu}\equiv \tfrac12\,\epsilon^{\mu\nu\alpha\beta}F_{\alpha\beta}\) the Hodge dual of \(F_{\mu\nu}\); in what follows we restrict to the parity-even case \(\mathcal{L}=\mathcal{L}(\mathcal{F})\).
For the specific model adopted below (following \cite{GhoshWalia2021}), the weak-field behavior is non-Maxwellian,
so the metric remains Schwarzschild-like at order \(1/r^2\) and the leading NED corrections enter at higher multipole order.

Varying the action \eqref{actionG} with respect to \(g_{\mu\nu}\) and \(A_\mu\) yields the field equations
\begin{equation}\label{EFE}
G_{\mu\nu}=8\pi\,T^{\text{NED}}_{\mu\nu},\qquad
\nabla_\mu\!\left(\mathcal{L}_{\mathcal{F}}\,F^{\mu\nu}\right)=0,\qquad
\nabla_\mu \tilde F^{\mu\nu}=0,
\end{equation}
where \(\mathcal{L}_{\mathcal{F}}\equiv \partial\mathcal{L}/\partial\mathcal{F}\).
The NED stress–energy tensor is
\begin{equation}\label{TNED}
T^{\text{NED}}_{\mu\nu}
=\frac{1}{4\pi}\Big(\mathcal{L}_{\mathcal{F}}\,F_{\mu\alpha}F_{\nu}{}^{\alpha}-g_{\mu\nu}\,\mathcal{L}(\mathcal{F})\Big).
\end{equation}
In the Maxwell limit \(\mathcal{L}=\mathcal{F}\), one recovers the standard electromagnetic stress–energy tensor within general relativity written as \(G_{\mu\nu}=8\pi\,T_{\mu\nu}\).

To construct the static seed solution, we adopt a Schwarzschild like gauge in which
\begin{equation}
\label{f0-seed}
ds^2=-f(r)\,dt^2+f(r)^{-1}dr^2+r^2\big(d\theta^2+\sin^2\theta\,d\phi^2\big),
\end{equation}
so that \(g_{tt}=-1/g_{rr}\). In the purely magnetic sector of NED we take the monopole ansatz
\begin{equation}
F_{\mu\nu}=2\,\delta^\theta_{[\mu}\delta^\phi_{\nu]}\,g\,\sin\theta\qquad\Longrightarrow\qquad 
F_{\theta\phi}=g\sin\theta,
\end{equation}
with magnetic charge \(g=\mathrm{const}\). The NED field equation \(\nabla_\mu(\mathcal{L}_{\mathcal{F}}F^{\mu\nu})=0\) then implies a constant monopole and yields the electromagnetic invariant
\begin{equation}
\mathcal{F}=\tfrac14 F_{\mu\nu}F^{\mu\nu}=\frac{g^2}{2r^4},
\end{equation}
see e.g. \cite{Ayon-Beato:2000mjt,Bronnikov2001}. The specific magnetically charged black–hole solution of interest is generated by choosing the NED Lagrangian density
\begin{equation}
\mathcal{L}(\mathcal{F})=\frac{2\,\sqrt{g}\;\mathcal{F}^{5/4}}{\,s\left(\sqrt{2}+2g\,\sqrt{\mathcal{F}}\right)^{3/2}},
\end{equation}
with \(s>0\) a constant that sets the nonlinearity scale \cite{GhoshWalia2021}. This static seed is then promoted to a rotating geometry via the revised Newman–Janis procedure (see, e.g., \cite{AzregAinou2014,GhoshWalia2021}).\\

Using the Lagrangian density in Eq.~\eqref{actionG} for the purely magnetic sector, the static seed solving the Einstein–NED system takes the Schwarzschild–like form
\begin{align}
\label{f-seed}
& ds^2=-f(r)\,dt^2+f(r)^{-1}dr^2+r^2(d\theta^2+\sin^2\theta\,d\phi^2), \\ &\text{with} \quad  f(r)=1-\frac{2M}{\sqrt{r^2+g^2}} \notag,
\end{align}
with ADM mass \(M\) and magnetic charge \(g\); in the limit \(g\to0\) the Schwarzschild metric is recovered \cite[Sec.~II, eq.~(8)]{GhoshWalia2021}.
In the monopole sector the NED field is
\(F_{\theta\phi}=g\sin\theta\), which implies a constant magnetic charge and the electromagnetic invariant
\begin{equation}
\label{F-invariant}
\mathcal{F}=\tfrac14 F_{\mu\nu}F^{\mu\nu}=\frac{g^2}{2r^4},
\end{equation}
as in the standard NED magnetic solutions \cite{Ayon-Beato:2000mjt,Bronnikov2001} and explicitly used in \cite[Sec.~II, eqs.~(5)–(7)]{GhoshWalia2021}.

The curvature invariants of the static seed metric \eqref{f-seed} provide a compact diagnostic of the geometry sourced by the NED monopole. Explicit expressions for the Ricci scalar and for the Kretschmann scalar \(K\equiv R_{\mu\nu\sigma\rho}R^{\mu\nu\sigma\rho}\) are reported in \cite[Sec.~II, below Eq.~(8)]{GhoshWalia2021}. For our purposes it is useful to recall their leading behaviours:
\begin{equation}
\begin{split}
K(r)&=\frac{48\,M^2}{r^6}+O(r^{-8})\qquad (r\to\infty),\\
K(r)&\propto \frac{M^2}{g^2}\,\frac{1}{r^{4}}\qquad (r\to 0).
\end{split}
\end{equation}
Thus, while the spacetime retains a scalar–polynomial singularity at the center, the NED scale \(g\) pushes the onset of large curvature inward and softens the central divergence with respect to the vacuum Schwarzschild case.
In other words, the metric function $f(r)$ remains finite (and vanishes at the horizon), while the curvature grows large only within a core $r\sim g$, as discussed in \cite{GhoshWalia2021}.

From a geometric–topological perspective, \(g\) plays the role of a magnetic charge for the two–form \(F\). In the purely magnetic sector one has
\begin{equation}
\frac{1}{4\pi}\int_{S^2}F = g,
\end{equation}
a flux that is conserved by the Bianchi identities and labels the cohomology class \([F]\) in the Wu–Yang/Dirac viewpoint. Within NED this charge sources a nonlinear stress–energy that (i) reduces to Maxwell electrodynamics at large radii and (ii) modifies the core geometry at \(r\!\sim\!g\), in line with the classical NED monopole constructions \cite{dirac,Ayon-Beato:2000mjt,Bronnikov2001}. 
Physically, introducing \(g\) is therefore not an arbitrary trick but a controlled way to encode strong–field electromagnetic self–interactions while preserving the correct weak–field limit.

\subsection{Weak gravitational lensing via the Gauss–Bonnet theorem}\label{lens}

Gravitational lensing is a sensitive probe of spacetime geometry and has been widely used to test black hole solutions in nonlinear electrodynamics (NED) by tracking how null rays are deflected \cite{Javed:2019kon,Javed:2020lsg,Okyay:2021nnh,Pantig:2022gih,Kumaran:2023brp,Waseem:2025yib} . In this section we employ the Gauss–Bonnet (GB) approach \cite{Gibbons:2008rj} to compute the weak deflection angle for the black hole metric introduced in Eq.~\eqref{f-seed}.

For null geodesics ($ds^2=0$) the optical metric is defined by
\begin{equation}
dt^2=\gamma_{ij}\,dx^i dx^j=\frac{1}{f^2(r)}\,dr^2+\frac{r^2}{f(r)}\,d\Omega^2, 
\label{opmetric}
\end{equation}
where $\gamma_{ij}$ is the spatial optical metric. Restricting to the equatorial plane ($\theta=\pi/2$), the problem reduces to the two–dimensional Riemannian manifold $(\mathcal{M}^{\rm opt},\gamma_{ab})$ with coordinates $(r,\phi)$.

The GB method requires the Gaussian curvature $\mathcal{K}$ of the optical surface. Writing $dS=\sqrt{\gamma}\,dr\,d\phi$ for the area element (with $\gamma\equiv\det\gamma_{ab}$) and using $R=2\mathcal{K}$ in two dimensions, we obtain to leading order
\begin{multline}
\mathcal{K}=\frac{R}{2}=-\frac{2 g^2 M^2}{\left(g^2+r^2\right)^3}+\frac{3 M^2 r^2}{\left(g^2+r^2\right)^3}+\\-\frac{g^2 M r^2}{\left(g^2+r^2\right)^{7/2}}-\frac{2 M r^4}{\left(g^2+r^2\right)^{7/2}}+\frac{g^4 M}{\left(g^2+r^2\right)^{7/2}}
.
\label{GC}
\end{multline}
Although Eq.~\eqref{GC} is exact, the weak-deflection result below is obtained by keeping only the \(O(M)\) contribution, consistent with the straight-line approximation \(r=b/\sin\phi\).

Let $D$ be a compact, oriented domain on $\mathcal{M}^{\rm opt}$ bounded by a piecewise smooth curve $\partial D$ with geodesic curvature $\kappa$ and exterior angles $\{\beta_i\}$. The Gauss–Bonnet theorem reads \cite{Gibbons:2008rj}
\begin{equation}
\iint_D \mathcal{K}\,dS+\oint_{\partial D}\kappa\,dt+\sum_{i}\beta_i=2\pi\chi(D),
\label{GB}
\end{equation}
where $\chi(D)$ is the Euler characteristic. Following Gibbons and Werner~\cite{Gibbons:2008rj}, we choose $\tilde D$ to be the region outside the light ray $C_1$ (a geodesic from source $S$ to observer $O$) and inside a circular arc $C_R:\,r(\phi)=R=\text{const}$ intersecting $C_1$ orthogonally at $S$ and $O$. For this construction $\kappa(C_1)=0$ and $\chi(\tilde D)=1$, so
\begin{equation}
\iint_{\tilde D}\mathcal{K}\,dS+\int_{C_R}\kappa(C_R)\,dt=\pi.
\label{GB2}
\end{equation}
With $\Gamma^{r}_{\phi\phi}$ the Christoffel symbol of \eqref{opmetric}, the nonzero geodesic–curvature piece along $C_R$ is
\begin{equation}
\kappa(C_R)=\!\left(\nabla_{\dot C_R}\dot C_R\right)^{\!r}
=\dot C_R^{\ \phi}\,(\partial_\phi \dot C_R^{\ r})+\Gamma^{r}_{\phi\phi}\,(\dot C_R^{\ \phi})^2,
\end{equation}
and in the limit $R\to\infty$ one finds
\begin{equation}
\lim_{R\to\infty}\big[\kappa(C_R)\,dt\big]=d\phi.
\label{geoR}
\end{equation}
Substituting \eqref{geoR} into \eqref{GB2} and noting that the total change of the polar angle along $C_R$ is $\int_{C_R} d\phi= \pi+\Theta$, we arrive at
\begin{equation}
\iint_{\tilde D}\mathcal{K}\,dS+\int_{0}^{\pi+\Theta} d\phi=\pi.
\end{equation}
Hence the weak deflection angle is \cite{Gibbons:2008rj}
\begin{equation}
\Theta=-\iint_{\tilde D}\mathcal{K}\,dS
= -\int_{0}^{\pi}\!\!\int_{b/\sin\phi}^{\infty}\! \mathcal{K}\,\sqrt{\gamma}\,dr\,d\phi,
\end{equation}
where we used the zeroth–order straight–line trajectory $r(\phi)=b/\sin\phi$ (impact parameter $b$). Evaluating the integral with the curvature \eqref{GC}, we obtain
\begin{equation}
\Theta \simeq
\frac{4 M}{b}-\frac{8 M \,g^{2}}{3 b^{3}}
\label{deflang}
\end{equation}

\begin{figure}[h]
    \centering \includegraphics[width=0.48\textwidth]{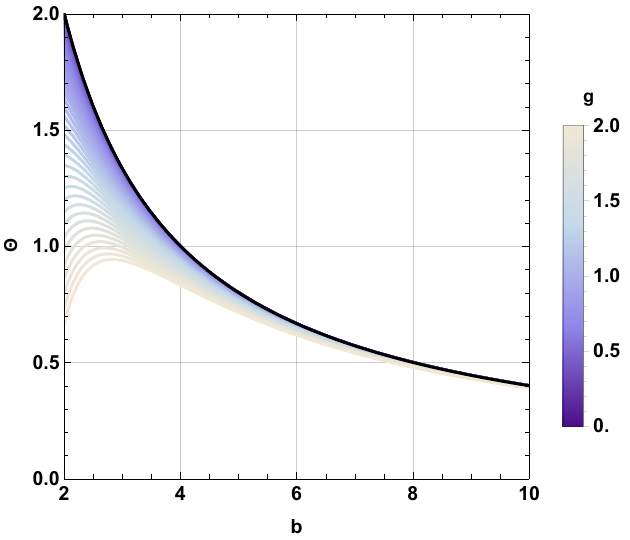}
    \caption{Figure shows the weak-field deflection angle $\Theta$ for the NED black hole with $M=1$. The black line corresponds to the Schwarzschild case $(g=0)$. 
    The horizontal axis is the impact parameter $b$, the vertical axis is the deflection angle $\Theta$, and the color bar encodes the NED parameter $g$. Larger $g$ suppresses bending at fixed $b$, while $g\!\to\!0$ recovers the Schwarzschild limit. Valid in the weak-field regime $b\gg M$.
 }
    \label{fig:WD1}
\end{figure}

The $g\to 0$ limit reproduces the purely GR terms, while the NED parameter $g$ generates model–dependent corrections beginning at $O(b^{-3})$. The dependence of the weak-field deflection angle on the impact parameter $b$ for different values of the NED parameter $g$ is shown in Fig.~\ref{fig:WD1}.

The constant \(g\) encodes the strength/scale of the NED sector sourcing the geometry (for instance, it coincides with the NED charge/scale in regular black–hole models). Its impact on lensing is that the leading NED correction, $-\tfrac{8 g^2 M}{3 b^3}$, \emph{decreases} the deflection for fixed $(M,b)$; thus $\partial\Theta/\partial(g^2)<0$ at $O(b^{-3})$. In the $g\to 0$ limit all NED contributions vanish, returning the standard GR result.

In the weak–field, thin–lens regime ($b\gg M$) the light–deflection angle of a static black hole reads $\Theta_{\rm stat}(b)=\frac{4 M}{b}-\frac{8 g^2 M}{3 b^3}+\mathcal{O}(b^{-4})$ (plus any model–dependent higher–order terms). For the slowly rotating (Kerr–like) case, to linear order in the spin $a$ and on the equatorial plane, the result is unchanged except for the well–known gravitomagnetic correction
\begin{equation}
\Theta(b)\simeq \Theta_{\rm stat}(b)\ \pm\ \frac{4aM}{b^{2}}+\mathcal{O}(a^{2}),
\end{equation}
where the upper (lower) sign corresponds to retrograde (prograde) trajectories. Thus rotation shifts the bending by $\pm 4aM/b^2$ without altering the static contributions at this order.

The Einstein ring is an observational phenomenon in gravitational lensing that occurs when a distant source, a massive lensing object, and an observer are perfectly aligned. Its angular radius can be derived from the general lens equation. In the weak-field limit, the relationship between the true angular position of the source ($\beta$), the apparent angular position of the image ($\theta$), and the deflection angle ($\Theta$) is
\begin{equation}
    \beta = \theta - \frac{D_{LS}}{D_S} \Theta(b),
\end{equation}
where $D_L$ is the distance from the observer to the lens, $D_S$ is the distance from the observer to the source, and $D_{LS}$ is the distance between the lens and the source. The impact parameter $b$ is related to the image angle by $b = \theta D_L$.

For perfect alignment, the source is directly behind the lens from the observer's perspective, so $\beta = 0$. The image forms a ring with an angular radius we denote as $\theta_E$. The condition becomes
\begin{equation}
    \theta_E = \frac{D_{LS}}{D_S} \Theta(\theta_E D_L).
\end{equation}
We substitute the given deflection angle Eq. \eqref{deflang} to the equation above. Multiplying by $\theta_E$ and rearranging the terms gives an equation for the squared radius, $\theta_E^2$:

\begin{equation}
    \theta_E^2 = \frac{4M D_{LS}}{D_L D_S} - \frac{8Mg^2 D_{LS}}{3\theta_E^2 D_L^3 D_S}.
\end{equation}
Such an equation can be solved perturbatively. We first find the zeroth-order solution, $\theta_{E,0}^2$, by ignoring the small correction term involving $g^2$, as
\begin{equation}
    \theta_{E,0}^2 = \frac{4M D_{LS}}{D_L D_S},
\end{equation}
which is the standard expression for the Einstein ring radius squared for a Schwarzschild black hole. Now, we substitute this approximation back into the denominator of the small correction term:
\begin{equation}
    \theta_E^2 \approx \theta_{E,0}^2 - \frac{8Mg^2 D_{LS}}{3(\theta_{E,0}^2) D_L^3 D_S}.
\end{equation}
Substituting the expression for $\theta_{E,0}^2$ allows us to simplify the correction as
\begin{equation} \label{einsring}
\begin{split}
    \theta_E^2 \approx \frac{4M D_{LS}}{D_L D_S} - \frac{8Mg^2 D_{LS}}{3 \left( \frac{4M D_{LS}}{D_L D_S} \right) D_L^3 D_S} =\\=\underbrace{\frac{4M D_{LS}}{D_L D_S}}_{\text{Schwarzschild Term}} - \underbrace{\frac{2g^2}{3D_L^2}}_{\text{NED Correction}}.
\end{split}
\end{equation}
Note that this approximation assumes \(\frac{2g^2}{3D_L^2}\ll \theta_{E,0}^2\), equivalently \(g^2 \ll \frac{6M D_{LS} D_L}{D_S}\).
The derived formula for the Einstein ring radius given in Eq. \eqref{einsring} reveals how nonlinear electrodynamics modifies this key gravitational lensing observable. The first term is the standard result from general relativity for a non-rotating, uncharged black hole. It depends on the black hole's mass $M$ and the geometric arrangement of the source, lens, and observer. The second term represents the modification introduced by the NED background. Its most important feature is the negative sign. This implies that for a non-zero value of $g$, the radius of the Einstein ring is smaller than what would be predicted for a standard Schwarzschild black hole of the same mass. This reduction provides a distinct and potentially measurable signature of the underlying NED geometry. The effect scales with $g^2$, meaning it is independent of the sign of the magnetic charge. High-precision measurements of Einstein rings, such as those that may become possible with next-generation telescopes, could place observational constraints on the parameter $g$. A measured ring radius that is systematically smaller than predicted by the mass estimate of a lensing object could be interpreted as evidence for modifications to general relativity, such as those described by this NED model.

\subsection{Shadow of the Einstein–NED black hole and EHT bounds for Sgr A*}
\label{subsec:shadow_EINED}

We consider the static, spherically symmetric Einstein–NED solution given in \ref{sec:level1}. For any metric of the form \ref{f0-seed}, unstable circular photon orbits satisfy
\begin{equation}
r\,f'(r)-2f(r)=0.
\end{equation}
With \ref{sec:level1} this gives
\begin{equation}
2Mg^2+3Mr^2-(g^2+r^2)\sqrt{g^2+r^2}=0.
\end{equation}
Introducing the dimensionless variables \(y\equiv r/M\) and \(k\equiv g/M\), the photon-sphere radius \(r_{\rm ph}\) is determined by
\begin{equation}
\label{eq:ph_eq_compact}
\big(k^2+y^2\big)^{3}=\big(3y^2+2k^2\big)^{2}.
\end{equation}
The critical impact parameter (shadow radius) then reads
\begin{equation}
\label{eq:bshadow_def}
b_{\rm sh}=\frac{r_{\rm ph}}{\sqrt{f(r_{\rm ph})}}
=\frac{M\,y_{\rm ph}}{\sqrt{1-\dfrac{2}{\sqrt{k^2+y_{\rm ph}^2}}}}\!,
\end{equation}
and the corresponding angular diameter for a source at distance \(D\) is \(\theta_{\rm sh}=2\,b_{\rm sh}/D\).

\paragraph*{Small-\(g\) expansion (analytic).}
Solving \eqref{eq:ph_eq_compact} perturbatively for \(k\ll 1\) yields
\begin{align}
\frac{r_{\rm ph}}{M}&=3-\frac{5}{18}\frac{g^2}{M^2}-\frac{37}{1944}\frac{g^4}{M^4}+O\!\left(\frac{g^6}{M^6}\right),\\[2mm]
\frac{b_{\rm sh}}{M}&=3\sqrt{3}\left[1-\frac{1}{18}\frac{g^2}{M^2}-\frac{7}{1944}\frac{g^4}{M^4}+O\!\left(\frac{g^6}{M^6}\right)\right].
\label{eq:bshadow_series}
\end{align}
Hence the fractional deviation from the Schwarzschild shadow is
\begin{equation}
\label{eq:delta_series}
\delta\ \equiv\ \frac{b_{\rm sh}(g)}{b_{\rm sh}(0)}-1
= -\frac{1}{18}\frac{g^2}{M^2}-\frac{7}{1944}\frac{g^4}{M^4}+O\!\left(\frac{g^6}{M^6}\right)\!<0,
\end{equation}
i.e., the NLED scale \(g\) monotonically \emph{shrinks} the shadow at leading orders.

The EHT data for Sgr A* reveal a bright, thick ring with diameter
\(\theta_{\rm ring}=51.8\pm2.3~\mu{\rm as}\) (68\% credible interval). Moreover, after calibrating to GRMHD libraries and adopting tight mass–distance priors, the observed image size is consistent with Kerr/GR predictions to within \(\sim10\%\)  \cite{EHT2022,Vagnozzi:2022moj,Do2020}.
Motivated by this, we impose
\begin{equation}
|\delta|\ \lesssim\ \varepsilon_{\rm EHT},\qquad 
\varepsilon_{\rm EHT}\simeq 0.10,
\end{equation}
as a conservative, model-agnostic criterion on deviations of the shadow size. From the analytic series \eqref{eq:delta_series} one obtains the closed-form estimate
\begin{equation}
\label{eq:bound_series}
\frac{g}{M}\ \lesssim\ \sqrt{18\,\varepsilon_{\rm EHT}}
\ \simeq\ 1.34 \quad (\text{keeping the leading }g^2\text{ term}).
\end{equation}
A more precise bound follows by solving the exact photon-sphere equation \eqref{eq:ph_eq_compact}, evaluating \eqref{eq:bshadow_def}, and enforcing \(\delta=-0.10\):
\begin{equation}
\label{eq:bound_exact}
\ \displaystyle \frac{g}{M}\ \lesssim\ 1.26\  \qquad (\delta=-0.10\ \text{threshold, exact}).
\end{equation}
Equations \eqref{eq:bound_series}–\eqref{eq:bound_exact} summarize the current Sgr A* constraint on the Einstein–NED scale \(g\) under the conservative requirement that the (geometrically defined) shadow—and thus the GRMHD-calibrated ring size—deviates by no more than \(\sim10\%\) from the Kerr expectation. Tighter (looser) choices of \(\varepsilon_{\rm EHT}\) map directly into correspondingly tighter (looser) limits on \(g/M\).

The mapping between the theoretical shadow and the observed ring is calibrated in the EHT analyses using extensive GRMHD libraries; the \(\sim10\%\) figure summarizes that calibration and its uncertainties for Sgr A* \cite{EHT2022,Vagnozzi:2022moj}. Our bounds use only the \emph{size} information and are complementary to constraints from visibility-domain morphology and polarimetry. Incorporating the posterior on \(M/D\) directly (instead of a single \(\varepsilon_{\rm EHT}\)) is straightforward and would further refine \eqref{eq:bound_exact}.

\begin{figure}
    \centering
    \includegraphics[width=0.49\textwidth]{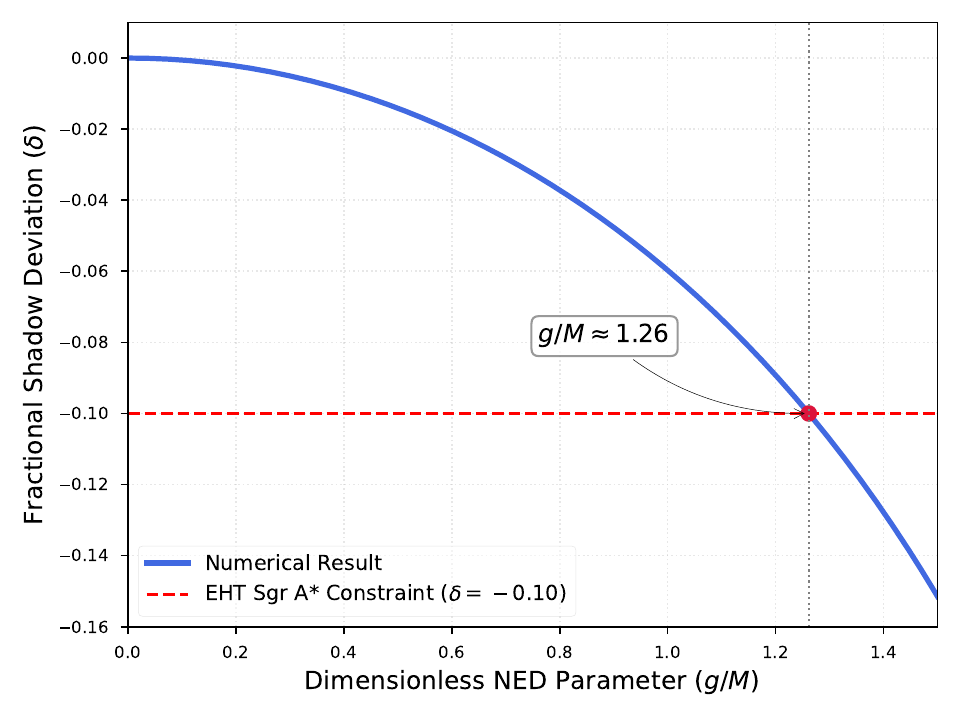}
    \caption{Constraint on the NED parameter from the shadow of Sgr A*. The plot shows the fractional deviation $\delta$ of the shadow radius from the Schwarzschild value as a function of the dimensionless NED parameter $g/M$. The solid curve is the exact numerical result. The horizontal dashed red line indicates the $\sim10\%$ observational constraint from EHT measurements ($\delta = -0.10$). The intersection point yields the upper bound $g/M \lesssim 1.26$, as derived in Eq.\eqref{eq:bound_exact}.}
    \label{fig:Shadow_Constraint}
\end{figure}

The relationship between the shadow deviation $\delta$ and the NED parameter $g/M$ is visualized in Figure \ref{fig:Shadow_Constraint}. The plot shows the numerically computed fractional deviation as a continuous curve, which monotonically decreases as $g/M$ increases. The EHT observational limit is represented by the horizontal dashed line at $\delta=-0.10$. The intersection of this line with the curve marks the upper bound on $g/M$, graphically confirming the result derived in Eq. \eqref{eq:bound_exact}. This visual representation shows how precision measurements of black hole shadows serve as a powerful tool to constrain or discover new physics beyond the standard paradigms of general relativity and Maxwell electrodynamics.

\subsection{Eikonal quasinormal modes of the Einstein–NED black hole}
\label{subsec:qnm_eikonal}

We compute the large–$\ell$ (eikonal) quasinormal modes (QNMs) of the static, spherically symmetric Einstein–NLED geometry
 \ref{f-seed}. To employ the eikonal formulae \cite{Churilova:2019jqx}, we match $f(r)$ to the large–$r$ expansion
\begin{equation}
f(r)=1-\frac{2M}{r}+\frac{\alpha_2}{r^2}+\frac{\alpha_3}{r^3}+\frac{\alpha_4}{r^4}+\frac{\alpha_5}{r^5}+\cdots .
\end{equation}
Our metric \ref{f-seed} reduces to this form
\begin{equation}
\label{eq:series_coeffs}
\begin{split}
& f(r)=1-\frac{2M}{r}+\frac{M g^2}{r^3}-\frac{3M g^4}{4 r^5}+\frac{5M g^6}{8 r^7}+\cdots ,\\
& \Rightarrow\quad
\alpha_2=0,\ \ \alpha_3=Mg^2,\ \ \alpha_4=0,\ \ \alpha_5=-\frac{3}{4}Mg^4.
\end{split}
\end{equation}

Eq.~\eqref{eq:series_coeffs} into Eq.~(20) yields, for multipole number $\ell\gg1$ and overtone $n=0,1,2,\dots$,
\begin{align}
\label{eq:qnm_eikonal_final}
\omega_{\ell n}
&=\frac{\ell+\frac12}{3\sqrt{3}\,M}
\left[1+\frac{\alpha_2}{6M^2}+\frac{\alpha_3}{18M^3}+\frac{\alpha_4}{54M^4}+\frac{\alpha_5}{162M^5}\right]\notag\\
&\quad-i\,\frac{n+\frac12}{3\sqrt{3}\,M}
\left[1+\frac{\alpha_2}{18M^2}-\frac{\alpha_3}{27M^3}-\frac{\alpha_4}{27M^4}-\frac{11\,\alpha_5}{486M^5}\right]\notag\\[1mm]
&=\frac{\ell+\frac12}{3\sqrt{3}\,M}
\left[1+\frac{g^2}{18M^2}-\frac{g^4}{216M^4}\right]\notag\\
&\quad-i\,\frac{n+\frac12}{3\sqrt{3}\,M}
\left[1-\frac{g^2}{27M^2}+\frac{11\,g^4}{648M^4}\right]
+O\!\Big(\tfrac{1}{\ell}\Big).
\end{align}

Equation~\eqref{eq:qnm_eikonal_final} may be written in the standard ``photon--ring'' form
\begin{equation}
\begin{split}
\Re\omega_{\ell n}=(\ell+\tfrac12)\,\Omega_c+O(\ell^0),\\
-\Im\omega_{\ell n}=(n+\tfrac12)\,\lambda_c+O(\ell^{-1}),
\end{split}
\end{equation}
with the orbital frequency and Lyapunov exponent
\begin{equation}
\begin{split}
\Omega_c=\frac{1}{3\sqrt{3}\,M}\left[1+\frac{g^2}{18M^2}-\frac{g^4}{216M^4}\right],\\
\lambda_c=\frac{1}{3\sqrt{3}\,M}\left[1-\frac{g^2}{27M^2}+\frac{11\,g^4}{648M^4}\right],
\end{split}
\end{equation}
valid through $O(g^4/M^4)$ within the linearized-$\alpha_i$ eikonal approximation.

The NED scale $g$ induces only even–power corrections, reflecting the $g\!\to\!-g$ symmetry of \eqref{f-seed}. To leading order, $g$ \emph{increases} the oscillation frequency (positive shift in $\Omega_c$) and \emph{decreases} the damping rate (smaller $\lambda_c$), hence longer–lived ringdown. The Schwarzschild result is recovered by setting $g\!=\!0$:
\(
\omega_{\ell n}^{\rm Schw}=\frac{\ell+\frac12}{3\sqrt{3}\,M}-i\,\frac{n+\frac12}{3\sqrt{3}\,M}+O(\ell^{-1}).
\)
Equation~\eqref{eq:qnm_eikonal_final} is accurate in the geometric–optics regime $\ell\gg1$ (test–field approximation on the fixed background); low–$\ell$ modes or strong–coupling regimes may require higher–order WKB or direct numerical integration for quantitative precision.

\vspace{10 pt}
We conclude this section by clarifying how we set the upper bound on the NED parameter $g$ used across this work. Our calibration is performed in the spherically symmetric sector (section~2.2) by matching the observed mean shadow diameter---EHT’s most robust scale observable. For completeness, we also report the corresponding eikonal QNM frequency shifts implied by the same $g/M$, as a consistency check. This choice is conservative once we move to the rotating Ghosh--Walia geometry of Eq.~\eqref{eq:NEDmetric}. In Kerr-like spacetimes the mean ring diameter depends only weakly (few per cent) on spin and viewing geometry, whereas rotation mainly affects the ring shape (asymmetry/circularity) rather than its overall size. Accordingly, the spherical ceiling $g/M \le 1.26$ safely overbounds the $g$ values allowed in the rotating setup and is adopted as fiducial throughout all $g$-scans in the main text.
Independent EHT analyses using the 2017 M87* data have placed shadow-size constraints on broad families of non-Kerr metrics and effective charges (see, e.g., \cite{Kocherlakota2021,Psaltis2020}), including several regular/NED-inspired or deformed spacetimes. To the best of our knowledge, the specific rotating Ghosh--Walia metric employed here was not explicitly covered in those surveys, which instead used model-agnostic parameterizations (e.g., KRZ; \cite{KonoplyaRezzollaZhidenko2016}) or other regular metrics. A recent study has also investigated the shadow-imaging properties of the same rotating NED geometry, including the dependence on spin, magnetic charge, and viewing angle using spherical light sources and thin accretion disks~\cite{YangGhoshKumar2025}. We therefore regard our spherical calibration as a robust, conservative benchmark for the rotating case; any future rotation-aware recalibration---yielding a spin-dependent $g_{\max}(a)$---would only tighten the constraint and thus strengthen our conclusions.


\section{\label{sec:level3} Rotating NED Black Hole}

To move from the spherical seed to an astrophysically relevant rotating geometry, we apply the revised Newman--Janis construction (without complexification) due to Azreg--A{\"i}nou~\cite{AzregAinou2014,Newman1965} to the metric~\eqref{f-seed}, following the implementation in~\cite[Sec.~II.1]{GhoshWalia2021}. This yields a Kerr--like line element depending on the mass parameter \(M\), the spin parameter \(a\), and the NED charge \(g\):
\begin{eqnarray}\label{eq:NEDmetric}
ds^2 &=& -\left(1 - \frac{2Mr^2}{\Sigma \sqrt{r^2 + g^2}}\right) dt^2
+ \frac{\Sigma}{\Delta} dr^2 + \Sigma d\theta^2 \nonumber\\
& & - \frac{4aMr^2}{\Sigma \sqrt{r^2 + g^2}} \sin^2 \theta \, dt \, d\phi
+ \frac{\mathcal{A}}{\Sigma} \sin^2 \theta \, d\phi^2,
\end{eqnarray}
where
\begin{equation}
\label{eq:NEDmetric1}
\Sigma = r^2 + a^2 \cos^2 \theta,
\end{equation}
\begin{equation}
\label{eq:NEDmetric2}
\Delta = r^2 + a^2 - \frac{2Mr^2}{\sqrt{r^2 + g^2}},
\end{equation}
\begin{equation}
\label{eq:NEDmetric3}
\mathcal{A} = (r^2 + a^2)^2 - a^2 \Delta \sin^2 \theta.
\end{equation}

This spacetime is singular at \(\Sigma=0\), corresponding to a ring-shaped curvature singularity. Horizons, when present, are located at the coordinate roots of \(\Delta(r)=0\), i.e.\ at null hypersurfaces of constant \(r\) determined by
\begin{equation}\label{eq:Delta=0}
r^2 + a^2 - \frac{2Mr^2}{\sqrt{r^2 + g^2}}=0\,.
\end{equation}
In what follows we denote by \(r_H\) the \emph{outer} horizon radius, i.e.\ the largest positive root of Eq.~\eqref{eq:Delta=0}.

\medskip

\begin{figure}[t]
    \centering
    \includegraphics[width=0.49\textwidth]{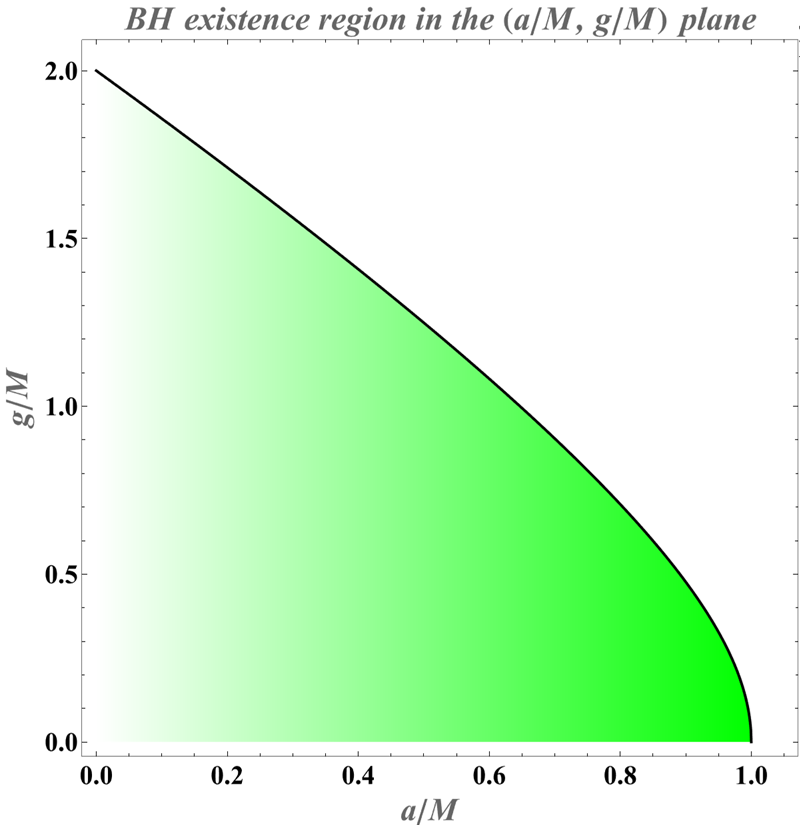}
    \caption{\textbf{Black-hole existence in the \((a/M,g/M)\) plane.}
    Shaded region: parameter values for which the horizon equation \(\Delta(r;a,g)=0\) admits at least one positive root, equivalently \(\min_{r>0}\Delta\le 0\).
    Thick black curve: extremal boundary where \(\Delta=\partial_r\Delta=0\), given in closed form by Eq.~\eqref{eq:extremal-curve}.
    The endpoints \((1,0)\) and \((0,2)\) correspond respectively to the Kerr extremal limit and to the static NED extremal limit.}
    \label{fig:BHregion}
\end{figure}

\noindent
To identify, in parameter space, the configurations that admit an event horizon, it is convenient to work with the dimensionless ratios \((a/M,g/M)\) and examine the horizon function \(\Delta(r;a,g)\) in Eq.~\eqref{eq:Delta=0}. For fixed \((a,g)\) the spacetime describes a black hole if and only if \(\Delta(r;a,g)=0\) has at least one positive root, equivalently if the global minimum over \(r>0\) satisfies \(\min_{r>0}\Delta \le 0\). The extremal boundary separating black holes from horizonless cores is reached when the minimum touches zero,
\begin{equation}
\Delta(r_e;a,g)=0,\qquad \partial_r\Delta(r_e;a,g)=0,
\end{equation}
which, introducing the dimensionless parameter \(s\equiv \sqrt{r_e^{\,2}+g^2}/M\), yields the closed parametric curve
\begin{equation}
\frac{a}{M}=(2-s)\sqrt{s},\qquad \frac{g}{M}=s\sqrt{s-1},\qquad s\in[1,2],
\label{eq:extremal-curve}
\end{equation}
interpolating smoothly between the Kerr extremal point \((a/M,g/M)=(1,0)\) and the static NED extremal point \((0,2)\). Figure~\ref{fig:BHregion} shows the resulting domain in the \((a/M,g/M)\) plane: the shaded region corresponds to parameter values for which an event horizon exists, while the thick black curve is the extremal boundary~\eqref{eq:extremal-curve}.

\medskip

\noindent
The ergoregion is bounded externally by the static-limit surface (SLS), defined by \(g_{tt}=0\), i.e.
\begin{equation}
(r^2 + a^2 \cos^2\theta)\sqrt{r^2+g^2}-2Mr^2=0.
\label{eq:SLS_general}
\end{equation}
On the equatorial plane \((\theta=\pi/2)\), we denote by \(r_L\) the corresponding static-limit radius (the equatorial SLS). For each \((a,g)\) admitting an event horizon, \(r_H\) and \(r_L\) are obtained by solving Eqs.~\eqref{eq:Delta=0} and~\eqref{eq:SLS_general} (at \(\theta=\pi/2\)), respectively. These two radii delimit the equatorial ergoregion interval \([r_H,r_L]\), which will be used in Sec.~\ref{sec:level4} as the integration domain for reconnection-driven extraction.

\medskip

\noindent
Further geometric diagnostics of the rotating NED spacetime---including the equatorial behavior of the outer horizon \(r_H\), the static-limit radius \(r_L\), the corresponding ergoregion thickness, as well as the photon circular orbits and ISCO structure---are reported in Appendix~\ref{app:sec3_supplement}. Since these quantities are not required explicitly in the extraction formalism developed in Sec.~\ref{sec:level4}, we keep them there as supplementary material supporting the geometric interpretation of the rotating background.

\section[Energy Extraction via the Comisso--Asenjo Process]{Energy Extraction via the Comisso--Asenjo Process in a Rotating NED Black Hole}
\label{sec:level4}

We investigate energy extraction via magnetic reconnection in a rotating black hole sourced by nonlinear electrodynamics (NED), working in the locally non-rotating (ZAMO) frame associated with the spacetime metric in Eqs.~\eqref{eq:NEDmetric}–\eqref{eq:NEDmetric3}. 

Physically, magnetic reconnection in a thin diffusion layer changes field-line connectivity and releases magnetic energy, launching two oppositely directed plasma outflows~\cite{PriestForbes2000,YamadaKulsrudJi2010,Zweibel2009}.
When the layer lies inside the ergoregion of a rotating spacetime, frame dragging allows one exhaust to carry negative Killing energy (as measured at infinity) and fall into the hole, while the opposite stream escapes with a surplus~\cite{Penrose1969,ComissoAsenjoPRD2021}. This is the fluid analogue of the Penrose mechanism: reconnection acts collectively on magnetofluid elements and, for upstream magnetizations \(\sigma_0\gtrsim1\), its efficiency is expected to be only weakly sensitive to microphysical details at the order-of-magnitude level~\cite{ComissoAsenjoPRD2021}. In disk-fed systems, magnetic-flux draping and azimuthal shear tend to form current sheets near the equatorial plane; we therefore start from a single equatorial sheet as a favorable baseline and assess how the background geometry modulates the negative-energy window and the overall extraction efficiency.

Within the small diffusion region we model the reconnecting magnetospheric field using the standard relativistic (ideal) MHD/Maxwell closure adopted in the Comisso--Asenjo framework; consequently, the NED parameter affects extraction mainly through the background geometry (redshift, frame dragging, and ergoregion extent), while the reconnection microphysics follows the rotating-spacetime analysis of~\cite{ComissoAsenjoPRD2021} and related literature. Operationally, this corresponds to treating the reconnecting field as a test electromagnetic field on the fixed NED-deformed geometry, i.e.\ neglecting its backreaction on the metric on the scales relevant to the diffusion region. 

To parametrize the strength of the NED sector sourcing the background solution, we introduce a characteristic invariant scale \(\Lambda_{\rm NED}\) (defined so that the electromagnetic invariants associated with the background NED configuration are of order \(\Lambda_{\rm NED}^4\) in the strong-field region). We then assume that the reconnecting magnetospheric field lies well below this scale, namely
\(
|F_{\mu\nu}F^{\mu\nu}|,\,|F_{\mu\nu}\tilde F^{\mu\nu}|\ll \Lambda_{\rm NED}^4,
\)
so that the local dynamics in the diffusion region can be consistently modeled with the standard RMHD/Maxwell constitutive relations used in the CA analysis.

Having framed the physical mechanism and working assumptions, we now introduce the ZAMO scalars that control redshift and frame–dragging in the rotating NED background.
To fix the geometric setting, we adopt geometrized units ($G=c=1$) and measure distances in units of the mass parameter $M$. In Boyer–Lindquist–like coordinates $(t,r,\theta,\phi)$, the locally non-rotating observers (ZAMOs) are characterized by three scalars that follow directly from the metric: the lapse $\alpha$, which measures the gravitational redshift between ZAMO proper time and coordinate time; the frame–dragging angular velocity $\omega$, namely the angular velocity of ZAMOs with respect to static observers at infinity; and the azimuthal proper (cylindrical) radius $\varpi\equiv\sqrt{g_{\phi\phi}}$, so that a circle at fixed $(r,\theta)$ has circumference $2\pi\varpi$. 
They read
\begin{equation}
\label{eq:ZAMOdefs}
\begin{aligned}
\alpha &= \bigl(-g^{tt}\bigr)^{-1/2}
= \sqrt{\frac{g_{t\phi}^2-g_{tt}\,g_{\phi\phi}}{\,g_{\phi\phi}\,}},\\
\omega &= -\frac{g_{t\phi}}{g_{\phi\phi}},\\
\varpi &= \sqrt{g_{\phi\phi}}\,.
\end{aligned}
\end{equation}
to be evaluated on Eqs.~\eqref{eq:NEDmetric}–\eqref{eq:NEDmetric3}. The ZAMO tetrad is orthonormal, hence the line element is locally Minkowskian,
\begin{equation}
ds^2=\eta_{\mu\nu}\,d\hat{x}^\mu d\hat{x}^\nu
=-d\hat{t}^{\,2}+d\hat{x}^{\,2}+d\hat{y}^{\,2}+d\hat{z}^{\,2},
\end{equation}
and local energies, momenta, and velocities are measured in this basis~\cite{Bardeen1972,ThornePriceMacdonald1986,Gourgoulhon2012}.

In a stationary, axisymmetric spacetime two Killing symmetries control particle and fluid dynamics: time translations generated by $k^\mu=(\partial_t)^\mu$ and axial rotations generated by $m^\mu=(\partial_\phi)^\mu$.  The associated integrals of motion are the Killing energy $\mathcal{E}=-p_\mu k^\mu$ and the axial angular momentum $L_z=p_\mu m^\mu$ (we use $L_z$ to avoid clashes with a length scale $\ell$). Their link to locally measured quantities follows from the standard $3{+}1$ decomposition in the ZAMO frame introduced above: the ZAMO energy $\hat{\mathcal{E}}$ and the conserved energy at infinity are related by
\begin{equation}
\hat{\mathcal{E}}=\frac{\mathcal{E}-\omega\,L_z}{\alpha}
\quad\Longleftrightarrow\quad
\mathcal{E}=\alpha\,\hat{\mathcal{E}}+\omega\,L_z,
\label{eq:ErelationZAMO}
\end{equation}
where $\alpha$ (lapse) and $\omega$ (frame–dragging angular velocity) are the geometric scalars defined in Eq.~\eqref{eq:ZAMOdefs}. Physically, $\alpha\,\hat{\mathcal{E}}$ is the redshifted local energy, while $\omega\,L_z$ accounts for rotational work due to frame dragging. Inside the ergoregion the stationary Killing vector becomes spacelike ($g_{tt}>0$), static observers do not exist, and $\mathcal{E}$ can be negative when the $\omega\,l_z$ term dominates the redshifted part—a geometric possibility underpinning Penrose splitting, wave superradiance, and, in our case, reconnection–driven outflows~\cite{Bardeen1972,ThornePriceMacdonald1986,Gourgoulhon2012,Penrose1969,ComissoAsenjoPRD2021}.

Restricting to the equatorial plane ($\theta=\pi/2$), we use the standard ZAMO orthonormal frame, in which locally measured azimuthal velocities take the form
$\hat v_\phi = (\sqrt{g_{\phi\phi}}/\alpha)\,(\Omega-\omega)$.
With our conventions (as in Ref.~\cite{ComissoAsenjoPRD2021}), the only nonzero shift component in the \emph{orthonormal} basis is
\begin{equation}
\beta^\phi \equiv \frac{\sqrt{g_{\phi\phi}}}{\alpha}\,\omega,
\label{eq:betaphiZAMO}
\end{equation}
so that $\hat v_\phi = (\sqrt{g_{\phi\phi}}/\alpha)\,\Omega - \beta^\phi$.
This should not be confused with the coordinate-basis shift, often written as $\beta^\phi_{\rm coord}=-\omega$; physically, $\beta^\phi$ is the frame-dragging contribution to the local azimuthal velocity.

The Keplerian angular velocity of circular equatorial geodesics in a stationary, axisymmetric spacetime reads
\begin{equation}
\Omega_{K\pm}=\frac{-\partial_r g_{t\phi}\ \pm\ \sqrt{\big(\partial_r g_{t\phi}\big)^2-\big(\partial_r g_{tt}\big)\big(\partial_r g_{\phi\phi}\big)}}{\ \partial_r g_{\phi\phi}\,},
\label{eq:OmegaK_general}
\end{equation}
(standard result, used in ZAMO-based analyses; see, e.g., \cite{ComissoAsenjoPRD2021} and references therein). The two roots correspond to equatorial circular orbits that are co- and counter-rotating with respect to the axial Killing direction \((\partial_\phi)^\mu\). Throughout this work we take \(a>0\) to fix the spin orientation and define the \emph{prograde} (co-rotating) branch as the one with \(\mathrm{sign}(\Omega_K)=\mathrm{sign}(a)\); equivalently, it is the branch that reduces in the Kerr limit to \(\Omega^{\rm Kerr}_{+}=+\sqrt{M}/(r^{3/2}+a\sqrt{M})\) and approaches \(+\sqrt{M}/r^{3/2}\) as \(r\to\infty\). With this convention, for $a>0$ the prograde orbit is $\Omega_{K+}$ (upper sign), while $\Omega_{K-}$ is retrograde.

For the rotating NED metric the general expression reduces to the closed form
\begin{equation}
\Omega_{K\pm}=\frac{\pm\,\sqrt{M}}{\,\bigl(r^{2}+g^{2}\bigr)^{3/4}\ \pm\ a\sqrt{M}\,},
\end{equation}
so that for \(a>0\) one has \(\Omega_{K+}>0\) (prograde) and \(\Omega_{K-}<0\) (retrograde). In the Kerr limit \(g\to0\) these expressions reduce to the familiar Kerr formulas \cite{ComissoAsenjoPRD2021}. In our plots and in the upstream kinematics entering Eqs.~\eqref{eq:vphi_comp}--\eqref{eq:gamma_comp} we adopt the prograde branch, i.e.\ \(\Omega_K\equiv\Omega_{K+}\), \(\hat v_K\equiv \hat v_{K+}\), and \(\hat\gamma_K\equiv \hat\gamma_{K+}\).

Finally, the Keplerian speed measured by ZAMOs follows directly from the mapping
\begin{equation}
\hat v_{K\pm}=\frac{\sqrt{g_{\phi\phi}}}{\alpha}\,\Omega_{K\pm}-\beta^\phi
=\frac{\sqrt{g_{\phi\phi}}}{\alpha}\,\big(\Omega_{K\pm}-\omega\big),
\end{equation}
and, after substituting the explicit expressions for $\alpha$, $\omega$ and $\Omega_{K\pm}$, one obtains the closed form
\begin{equation}
\label{eq:vKpm_compact}
\begin{aligned}
\hat v_{K\pm}
&= -\,\frac{\sqrt{M}\,\sqrt{\,r^2+a^2\!\left(1+\frac{2M}{s}\right)}}{\,s^3-a^2M\,}
\\
&\quad\times\Bigg[
\pm\,s^{3/2}+\frac{a\sqrt{M}}{s}\Big(-2a^2M+s\big(1+2s^2\big)\Big)
\Bigg]
\\
&\quad\times \sqrt{\frac{r^2s^2+a^2\!\left(s^2+2Ms\right)}{a^2s^2+r^2\,s\left(s-2M\right)}}\,,
\qquad s\equiv\sqrt{r^2+g^2}\,.
\end{aligned}
\end{equation}

Having established the ZAMO kinematics, we now set up the plasma model and the energy-at-infinity conditions needed for extraction. 
We model the plasma in the ergoregion with the single–fluid energy–momentum tensor used by Comisso and Asenjo,
\begin{equation}
T^{\mu\nu}
= p\,g^{\mu\nu}+w\,U^\mu U^\nu
+F^\mu{}_{\delta}F^{\nu\delta}
-\tfrac{1}{4}\,g^{\mu\nu}F^{\rho\delta}F_{\rho\delta},
\label{eq:Tmunu_onefluid}
\end{equation}
where $p$ is the pressure, $w$ the enthalpy density (including rest–mass and thermal parts), $U^\mu$ the fluid four–velocity, and $F^{\mu\nu}$ the electromagnetic field tensor~\cite{ComissoAsenjoPRD2021}. The quantity that governs extraction is the energy density at infinity associated with the timelike Killing field,
\begin{equation}
e^\infty \;\equiv\; -\,\alpha\,g^{\mu 0}\,T_{\mu 0},
\label{eq:einf_def}
\end{equation}
which in the ZAMO frame can be written compactly as
\begin{equation}
e^\infty \;=\; \alpha\Big(\hat e + \beta^\phi \hat p^{\hat\phi}\Big),
\end{equation}
where \(\hat e \equiv T^{\hat 0\hat 0}\) is the ZAMO-measured energy density and \(\hat p^{\hat\phi}\equiv T^{\hat 0\hat\phi}\) is the ZAMO-measured azimuthal momentum density. Within the one-fluid model described by tensor \eqref{eq:Tmunu_onefluid}, the energy density at infinity  \eqref{eq:einf_def} naturally decomposes into a hydrodynamic and an electromagnetic part,
\begin{equation}
\begin{split}
e^\infty
&=
\underbrace{\alpha\big(w\hat\gamma^{2}-p\big)
+ \alpha\,\beta^\phi\,w\,\hat\gamma^{2}\hat v_\phi}_{\displaystyle e^\infty_{\rm hyd}}\\
&\quad+\;
\underbrace{\alpha\,\tfrac{\hat B^{2}+\hat E^{2}}{2}
+ \alpha\,\beta^\phi\,(\hat{\boldsymbol B}\!\times\!\hat{\boldsymbol E})_\phi}_{\displaystyle e^\infty_{\rm em}},
\end{split}
\label{eq:einf_split}
\end{equation}
where hats denote ZAMO–measured quantities, $\hat v_\phi$ is the azimuthal component of the fluid 3–velocity in that frame, and $\hat\gamma=(1-\hat v^{2})^{-1/2}$ is the associated Lorentz factor. Equation~\eqref{eq:einf_split} makes explicit the roles of gravitational redshift ($\alpha$) and frame dragging (terms $\propto\beta^\phi$) in the energy budget~\cite{ComissoAsenjoPRD2021}.

In the reconnection diffusion region---the X-line, i.e., the short segment of magnetic nulls where field lines change connectivity and the plasma decouples from the magnetic field~\cite{PriestForbes2000,YamadaKulsrudJi2010}---we assume that, over the layer-crossing time, the dissipated magnetic energy is efficiently converted into particle bulk/thermal energy and that the fluid element evolves approximately adiabatically. Under this assumption, the electromagnetic contribution from the immediate downstream is subleading relative to the hydrodynamic part that reaches infinity~\cite{ComissoAsenjoPRD2021}. Relativistic particle--in--cell (PIC) simulations of collisionless reconnection in magnetically dominated plasmas (\(\sigma_0\gtrsim1\)) typically find that a substantial fraction (\(\sim50\)–\(80\%\)) of the released magnetic energy is transferred to particle bulk/thermal energy within a few layer thicknesses downstream of the X-line; for \(\sigma_0\gg1\) the converted fraction can be even larger~\cite{SironiSpitkovsky2014,Guo2015,WernerUzdensky2016,GuoReview2020}.
Under this assumption we retain only the hydrodynamic contribution,
\begin{equation}
e^\infty\approx e^\infty_{\rm hyd}
=\alpha\Big[(w\,\hat\gamma^{2}-p)+\beta^\phi\,w\,\hat\gamma^{2}\,\hat v_\phi\Big]\,,
\label{eq:einf_hyd}
\end{equation}
which captures the \emph{sign} and the leading magnitude of the near--layer budget.

For completeness, the electromagnetic remainder can be bounded directly from Eq.~\eqref{eq:einf_split}. Using the inequality
$|(\hat{\boldsymbol B}\times\hat{\boldsymbol E})_\phi|\le (\hat B^2+\hat E^2)/2$, one obtains
\begin{equation}
|e^\infty_{\rm em}|
=\alpha\left[\frac{\hat B^2+\hat E^2}{2}+\beta^\phi(\hat{\boldsymbol B}\times\hat{\boldsymbol E})_\phi\right]
\le \alpha\,\frac{\hat B^2+\hat E^2}{2}\,\bigl(1+|\beta^\phi|\bigr)\,.
\label{eq:einf_em_bound}
\end{equation}
PIC conversion fractions motivate the parametrization of the residual field energy density in the immediate downstream as
$(\hat B^2+\hat E^2)_{\rm ds}/2 \equiv f_{\rm em}\,b_0^2/2$ with $0\le f_{\rm em}\lesssim 0.2$--$0.5$ for $\sigma_0\gtrsim1$ (particle-dominated downstream). Here \(f_{\rm em}\) is meant as the small residual fraction after the rapid conversion stage; using it in the estimate below is therefore consistent with the efficient-conversion closure employed later. We use these PIC-motivated fractions only at the order-of-magnitude level; frame-dependent differences between comoving and ZAMO measures on the layer scale amount to factors of order unity.
With $b_0^2=\sigma_0 w_0$ and the closure $w=(1+\sigma_0)w_0$ adopted below to derive Eq.~\eqref{eq:epsiloninfty}, Eqs.~\eqref{eq:einf_hyd} and~\eqref{eq:einf_em_bound} yield the conservative estimate
\begin{equation}
\frac{|e^\infty_{\rm em}|}{|e^\infty_{\rm hyd}|}
\;\lesssim\;
\frac{(1+|\beta^\phi|)\,f_{\rm em}\,\sigma_0}
{2(1+\sigma_0)\left|\hat\gamma^{2}(1+\beta^\phi \hat v_\phi)-p/w\right|}\,.
\label{eq:einf_ratio_bound}
\end{equation}
For the relativistically hot closure used later (\(\Gamma=4/3\), hence \(p/w\simeq 1/4\)) and for the parameter ranges explored in our grids, Eq.~\eqref{eq:einf_ratio_bound} implies that the electromagnetic remainder is generically a subleading correction to the near-layer energy-at-infinity budget whenever $f_{\rm em}\lesssim\mathcal{O}(1)$, and in particular it is bounded at the level of \(\lesssim\) few \(\times 10^{-1}\) for $f_{\rm em}\lesssim 0.5$ and for the kinematic factors relevant to fast reconnection. 
Therefore, neglecting $e^\infty_{\rm em}$ mainly affects the normalization of the extracted power, while the sign diagnostics defining the extraction criteria remain stable except possibly in a narrow neighborhood of threshold, which is excluded by the extraction mask $f_{\rm extr}(r)$ (and, in practice, by the detectability floor used in the scans). We stress that this near--layer approximation does not preclude Poynting--dominated transport on larger scales; it only concerns the local bookkeeping across the diffusion region, as in Ref.~\cite{ComissoAsenjoPRD2021}.

The upstream plasma rotates azimuthally with ZAMO speed $\hat v_K$ and Lorentz factor $\hat\gamma_K$. In the upstream comoving frame $x^{\mu'}$~\cite{ComissoAsenjoPRD2021}, reconnection launches two exhausts with speed $v_{\rm out}$ whose direction is parametrized by the \emph{outflow pitch angle} $\xi$, defined by $\tan\xi \equiv v^{1'}_{\rm out}/v^{3'}_{\rm out}$ (so that $\xi=0$ is purely azimuthal and $\xi=\pi/2$ purely radial).
Only the azimuthal projection enters the frame--dragging coupling, hence $\cos\xi$ appears in the ZAMO azimuthal velocities obtained by relativistic velocity addition:
\begin{equation}
\hat v_\phi^{\,\pm}
=\frac{\hat v_K \pm v_{\rm out}\cos\xi}{1\pm \hat v_K v_{\rm out}\cos\xi},
\label{eq:vphi_comp}
\end{equation}
with corresponding Lorentz factors
\begin{equation}
\hat\gamma_\pm
=\hat\gamma_K\,\gamma_{\rm out}\,\big(1\pm \hat v_K v_{\rm out}\cos\xi\big).
\label{eq:gamma_comp}
\end{equation}

In what follows, we use “fast reconnection” in its quantitative sense: a plasmoid--mediated reconnection rate
\(\mathcal{R}\equiv v_{\rm in}/v_A\simeq 0.1\), only weakly sensitive to microphysics, as supported by relativistic PIC studies in magnetically dominated plasmas~\cite{SironiSpitkovsky2014,Guo2015,WernerUzdensky2016,GuoReview2020}.
Here \(v_{\rm in}\), \(v_A\), and \(v_{\rm out}\) are understood as measured in the local rest frame of the upstream plasma (as in Ref.~\cite{ComissoAsenjoPRD2021}).

The relativistic Alfv\'en speed \(v_A\) is defined in the fluid rest frame; under the ideal--MHD condition
\(E'^{\mu}\equiv F^{\mu\nu}U_{\nu}=0\) one finds
\begin{equation}
v_A^2 \;=\; \frac{b_0^2}{w_0+b_0^2} \;=\; \frac{\sigma_0}{1+\sigma_0}\qquad (c=1),
\label{eq:vA_def}
\end{equation}
where \(b_0^2\equiv b^\mu b_\mu\) is the upstream comoving magnetic--field invariant (primes/comoving frame), \(w_0\) is the upstream total enthalpy density,
and \(\sigma_0\equiv b_0^2/w_0\) is the upstream magnetization parameter~\cite{Anile1989,Lyubarsky2005}.
(When translating to Gaussian cgs, one may use \(b_0^2=B_0'^2/4\pi\) so that \(B_0'\) is in Gauss.)

In thin plasmoid--mediated layers the outflow speed approaches \(v_A\)~\cite{Lyubarsky2005,ComissoAsenjoPRD2021}; accordingly we adopt
\begin{equation}
v_{\rm out}\simeq v_A(\sigma_0),\qquad
\gamma_{\rm out}\simeq \frac{1}{\sqrt{1-v_{\rm out}^2}}\simeq \sqrt{\,1+\sigma_0\,},
\label{eq:reconn_scalings}
\end{equation}
as in Ref.~\cite{ComissoAsenjoPRD2021}.

The characteristic timescales of a reconnection layer of half--length \(L\) and thickness \(\delta_X\) are
\begin{equation}
t_{\rm rec}\sim\frac{L}{\mathcal{R}\,v_A},\qquad
t_{\rm cross}\sim\frac{\delta_X}{v_A}.
\end{equation}
For \(\sigma_0=3\) one has \(v_A=\sqrt{3/4}\simeq0.87\), hence \(t_{\rm rec}\approx 11.5\,L\) for \(\mathcal{R}\simeq0.1\) and \(t_{\rm cross}\approx1.15\,\delta_X\) (in units with \(c=1\)).
Taking \(L\approx5\,M\) near the horizon gives \(t_{\rm rec}\approx60\,M\).
Using \(1\,M=GM/c^3\simeq4.93\,\mu{\rm s}\,(M/M_\odot)\), this corresponds to \(\sim3\times10^{4}\) s (\(\sim8\) h) for \(10^{8}M_\odot\) and \(\sim3\times10^{-3}\) s for \(10\,M_\odot\).
Thus individual near--horizon reconnection episodes evolve on \(\mathcal{O}(10\!-\!10^2)\) gravitational times, while longer-timescale variability is expected to be regulated by the large-scale inflow and magnetic-flux supply.
At fixed geometry, increasing \(\sigma_0\) increases \(v_A\) (and \(\gamma_{\rm out}\)), which enhances the kinetic contribution in Eq.~\eqref{eq:einf_hyd} and tends to widen the parameter region where a negative-energy counter-rotating branch can occur.
Starting from the local energy density at infinity in the ZAMO frame, Eq.~\eqref{eq:einf_hyd}, and inserting (i) the kinematics of the co--/counter--rotating outflow via Eqs.~\eqref{eq:vphi_comp}--\eqref{eq:gamma_comp}, and (ii) the thermodynamic closure in the diffusion region (relativistically hot plasma, \(\Gamma=4/3\)), we express the downstream enthalpy as \(w=(1+\sigma_0)w_0\) (Maxwell/ideal limit), with the upstream magnetization \(\sigma_0=b_0^2/w_0\), to obtain the \emph{energy at infinity per upstream enthalpy} for the two branches (cf.\ the Kerr derivation in~\cite{ComissoAsenjoPRD2021}):
\begin{equation}
\label{eq:epsiloninfty}
\begin{split}
&\epsilon^\infty_\pm(r;\,a,g,\sigma_0,\xi)
= \alpha\,\hat{\gamma}_K \Bigg[
(1+\beta^\phi \hat{v}_K)\,\sqrt{1+\sigma_0} \\
&\;\pm\ \cos\xi\,(\beta^\phi+\hat{v}_K)\,\sqrt{\sigma_0}
-\ \frac{1}{4}\,
\frac{\sqrt{1+\sigma_0}\ \mp\ \cos\xi\,\hat{v}_K\,\sqrt{\sigma_0}}
{\hat{\gamma}_K^{2}\,\big(1+\sigma_0-\cos^{2}\xi\,\hat{v}_K^{2}\sigma_0\big)}
\Bigg]\,.
\end{split}
\end{equation}
Here \(\alpha=\alpha(r;a,g)\), \(\beta^\phi=\beta^\phi(r;a,g)\), \(\hat v_K=\hat v_K(r;a,g)\), and \(\hat\gamma_K=\hat\gamma_K(r;a,g)\) are evaluated on the equatorial plane, with definitions given in Eqs.~\eqref{eq:ZAMOdefs}, \eqref{eq:betaphiZAMO}, and in the Keplerian relations above.
With the branch energies at infinity in hand, the extraction window is fixed by the sign of the counter–rotating branch and by the positivity of the co–rotating one. The square bracket (overall multiplied by \(\alpha\,\hat\gamma_K\)) contains the redshifted enthalpy load and its modulation by frame dragging (\(\propto\beta^\phi\)) and orbital motion (\(\hat v_K\)). The final rational term encodes the adiabatic/pressure correction, most relevant at low magnetization or for strongly oblique outflows (large \(\xi\)). The \(\pm\) sign labels the co-/counter-rotating exhausts defined by the azimuthal velocities in Eqs.~\eqref{eq:vphi_comp}–\eqref{eq:gamma_comp}: inside the ergoregion, sufficiently large \((\beta^\phi+\hat v_K)\) and nearly azimuthal launch (\(\cos\xi\simeq1\)) can drive the counter-rotating branch (\(-\)) to negative Killing energy at infinity, while the co-rotating branch (\(+\)) emerges with a surplus~\cite{Penrose1969,ComissoAsenjoPRD2021}.

In the equatorial plane ($\theta=\pi/2$; axisymmetry removes any explicit $\phi$--dependence), black--hole energy extraction by reconnection occurs when the two standard conditions (cf.\ Eqs.~(34)--(36) of \cite{ComissoAsenjoPRD2021}) are met:
\begin{align}
&\epsilon^\infty_-(r;\,a,g,\sigma_0,\xi) \;<\; 0,
\label{eq:cond_neg}\\[4pt]
\Delta\epsilon^\infty_+(r;\,a,g,\sigma_0,\xi)
\;\equiv\;
&\epsilon^\infty_+(r;\,a,g,\sigma_0,\xi)
\nonumber\\
&\qquad-\;\Big(1-\frac{\Gamma}{\Gamma-1}\frac{p}{w}\Big)
\;>\; 0.
\label{eq:cond_pos}
\end{align}
The first inequality requires that the decelerated (counter-rotating) branch carry \emph{negative} Killing energy to the horizon. The second demands that the accelerated (co--rotating) branch emerge with energy at infinity larger than its rest--mass plus thermal contribution per unit enthalpy. For a relativistic plasma with adiabatic index \(\Gamma=4/3\) the comoving internal energy density is \(u=p/(\Gamma-1)=3p\), hence \(w=\rho+u+p=\rho+4p\) and in the ultra-relativistic regime in which the rest--mass contribution is subdominant (\(\rho\ll u\)), one obtains \(p/w\simeq1/4\). Consequently, the bracket in Eq.~\eqref{eq:cond_pos} vanishes and the condition reduces to \(\epsilon^\infty_+>0\). 

Taken together with Eq.~\eqref{eq:epsiloninfty}, the conditions \eqref{eq:cond_neg}–\eqref{eq:cond_pos} highlight the key geometric control parameters: the azimuthal coupling to frame dragging enters through \((\beta^\phi+\hat v_K)\) and is weighted by $\cos\xi$, where $\xi$ is the outflow pitch angle relative to the azimuthal direction defined in the upstream comoving frame~\cite{ComissoAsenjoPRD2021}. The dependence on \(\xi\) is strongly nonlinear: extraction is sharply enhanced for nearly azimuthal launch (\(\cos\xi\to1\)), while modest departures can substantially shrink or suppress the negative-energy window, especially at moderate \(\sigma_0\).

Once the sign of the energy at infinity is controlled (i.e., $\epsilon_-^\infty<0$ and $\epsilon_+^\infty>0$),
the total extracted power follows by integrating the negative-energy flux over the allowed region. Following Comisso--Asenjo's formulation (recast in ZAMO variables) we write
\begin{equation}
P_{\rm extr}(a,g;\sigma_0,\xi)\;=\;\int_{A_{\rm in}}\!\Big[-\,\epsilon^\infty_-(\mathbf{x};\,a,g,\sigma_0,\xi)\Big]\; w_0\,U_{\rm in}\; dA,
\label{eq:Pextr_def}
\end{equation}
where $w_0$ is the upstream enthalpy density, $U_{\rm in}$ the inflow speed into the layer, and $A_{\rm in}$ the portion of the equatorial plane where the extraction criteria \eqref{eq:cond_neg}--\eqref{eq:cond_pos} are fulfilled (cf.\ \cite{ComissoAsenjoPRD2021}). \\
In axisymmetry, restricting to a single equatorial current sheet, Eq.~\eqref{eq:Pextr_def} reduces to the radial form
\begin{equation}
\label{eq:Pextr_equatorial}
\begin{aligned}
&P_{\rm extr}(a,g;\sigma_0,\xi)
=2\pi\,w_0\,U_{\rm in}\!
\int_{r_H}^{r_L}
f_{\rm extr}(r)\,
\\
&\qquad \times
\Big[-\epsilon^\infty_-(r;\,a,g,\sigma_0,\xi)\Big]\,
\sqrt{\gamma_{rr}(r)\gamma_{\phi\phi}(r)}\,dr \,.
\end{aligned}
\end{equation}
with \(\epsilon^\infty_\pm(r;\,a,g,\sigma_0,\xi)\) given by Eq.~\eqref{eq:epsiloninfty}.

Here \(f_{\rm extr}(r)\in\{0,1\}\) selects the radii satisfying Eqs.~\eqref{eq:cond_neg}--\eqref{eq:cond_pos}.
The integration range \([r_H,r_L]\) spans the equatorial ergoregion between the outer horizon radius \(r_H\) and the equatorial static limit \(r_L\) (defined by \(g_{tt}(r_L)=0\)), while the measure in Eq.~\eqref{eq:Pextr_equatorial} is the proper area element of the equatorial sheet at fixed \(t\): \(dA=\sqrt{\det(\gamma_{AB})}\,dr\,d\phi\) with \(\gamma_{AB}=g_{AB}|_{\theta=\pi/2}\) (\(A,B=r,\phi\)). In Boyer--Lindquist--type coordinates \(g_{r\phi}=0\), hence \(dA=\sqrt{g_{rr}g_{\phi\phi}}\,dr\,d\phi\) and the azimuthal integration yields the prefactor \(2\pi\); for our metric \(\sqrt{g_{rr}g_{\phi\phi}}=\sqrt{\mathcal{A}(r)/\Delta(r)}\) at \(\theta=\pi/2\).

For scaling purposes, one may obtain a simple estimate by taking \(w_0\) and \(U_{\rm in}\) approximately constant and evaluating the integrand at a representative radius over a width \(\Delta r\simeq r_L-r_H\); the results shown below always use the full integral in Eq.~\eqref{eq:Pextr_equatorial}.
To isolate geometry and kinematics, it is convenient to define the enthalpy-normalized proxy
\begin{equation}
\label{eq:proxy_I}
\begin{aligned}
&\mathcal{I}(a/M,g/M;\sigma_0,\xi)\;\equiv \\[-2pt]
&\qquad \int_{r_H}^{r_L}\!
f_{\rm extr}(r)\,\Big[-\,\epsilon^\infty_-(r;\,a,g,\sigma_0,\xi)\Big]\;
\sqrt{\frac{\mathcal{A}(r)}{\Delta(r)}}\;dr\,.
\end{aligned}
\end{equation}

\begin{figure*}[t]
    \centering
    \subfigure{
        \includegraphics[height=7.5cm]{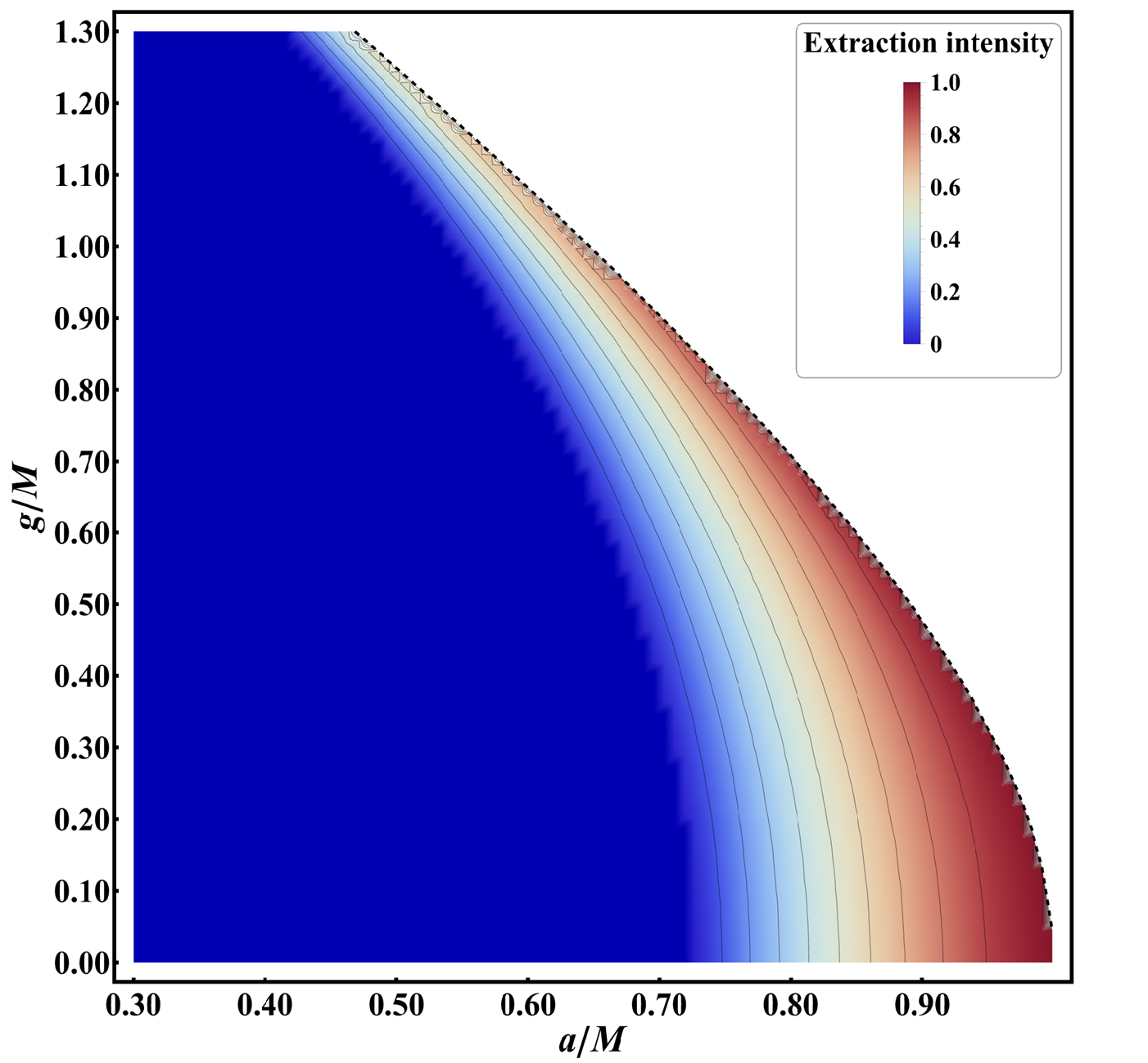}
    }\hfill
    \subfigure{
        \includegraphics[height=7.5cm,trim=170 0 0 0,clip]{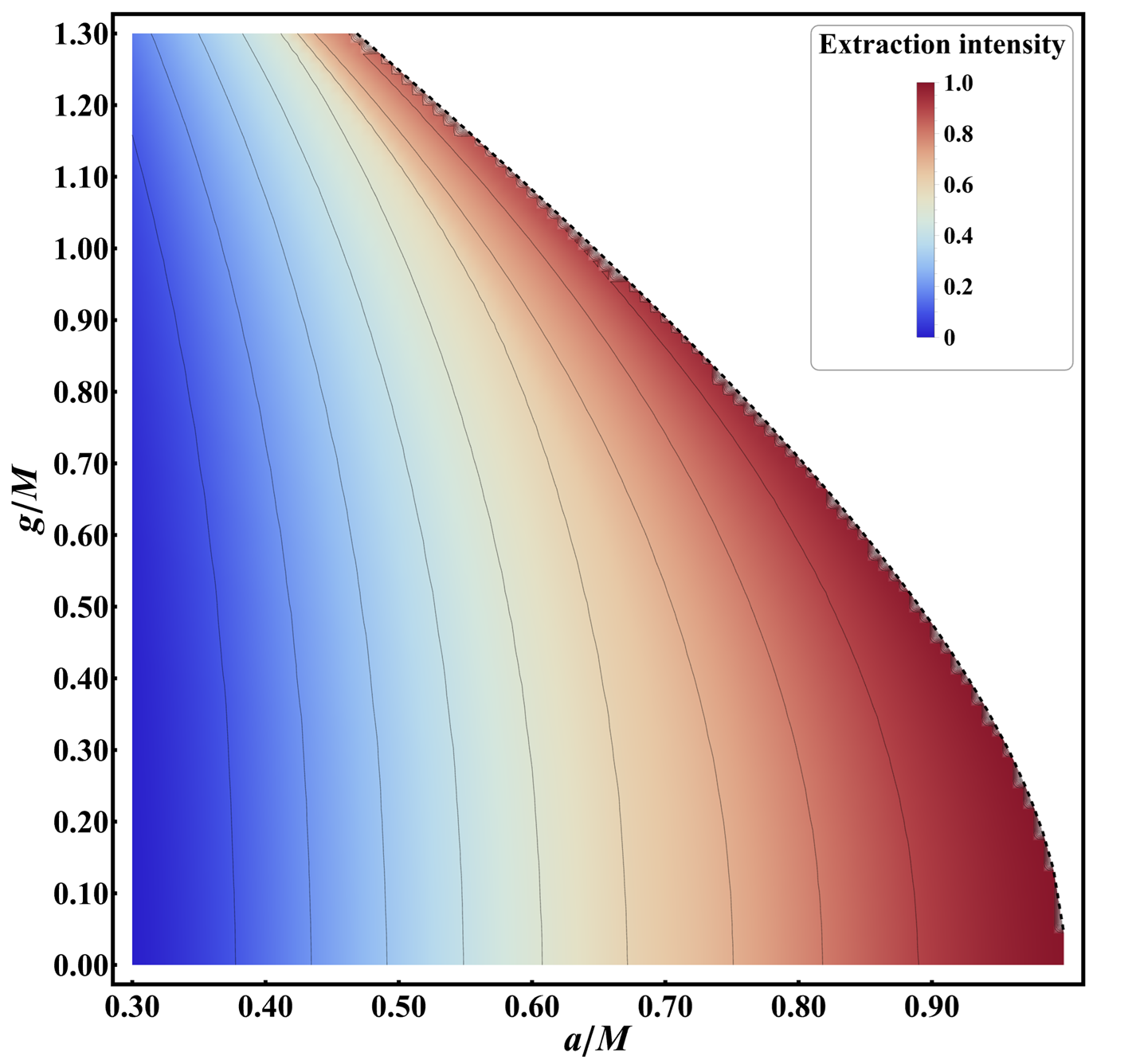}
    }\\[2mm]
    \subfigure{
        \includegraphics[height=7.5cm]{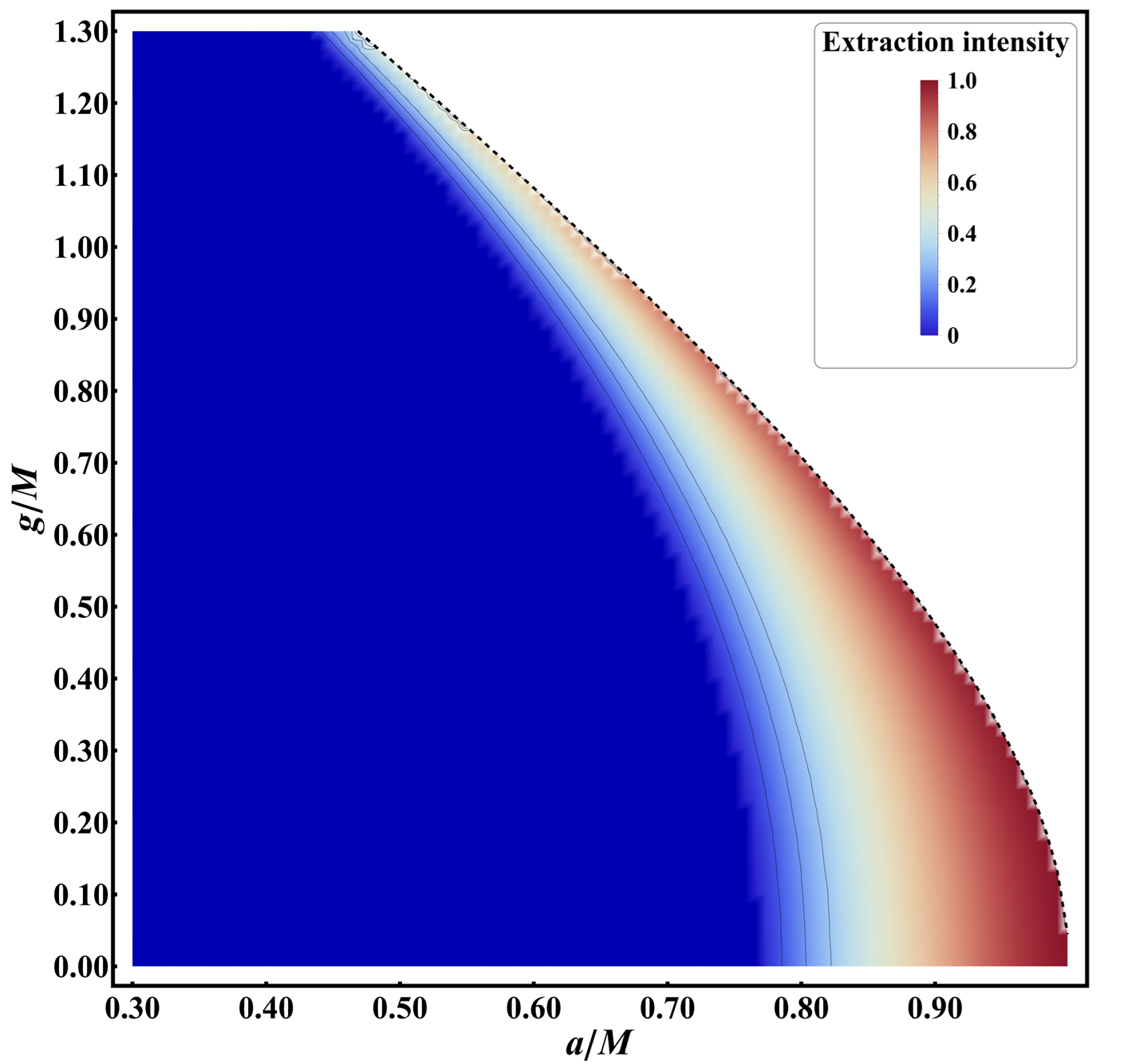}
    }\hfill
    \subfigure{
        \includegraphics[height=7.5cm,trim=170 0 0 0,clip]{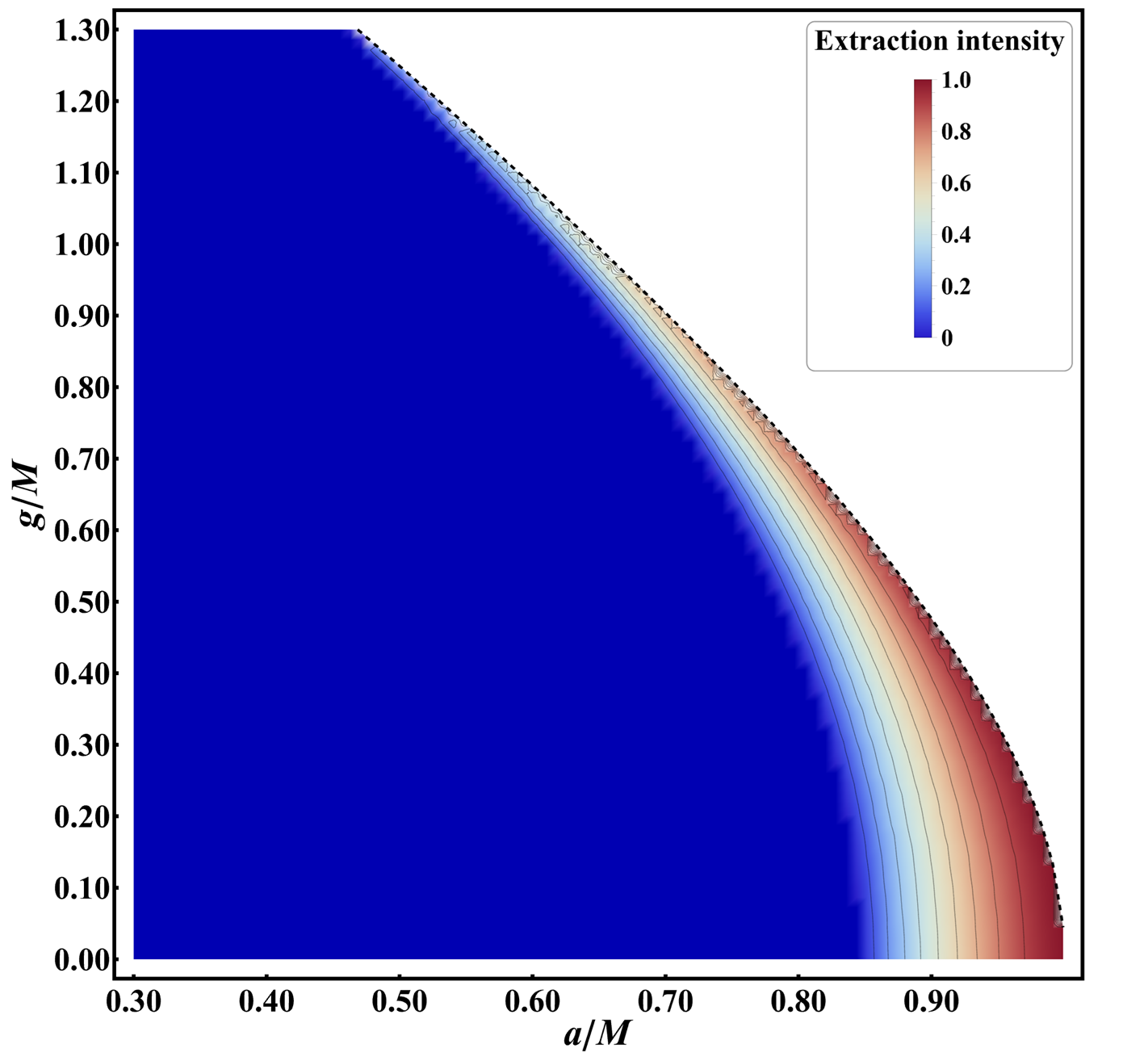}
    }
    \caption{\small\textbf{Extraction maps in the \((a/M,\,g/M)\) plane.}
Color shows the proxy \(\mathcal{I}(a/M,g/M;\sigma_0,\xi)\) defined in Eq.~\eqref{eq:proxy_I}, proportional to \(P_{\rm extr}\) up to the constant prefactor \(2\pi w_0U_{\rm in}\).
Top row: \(\xi=0\) with (a) \(\sigma_0=500\), (b) \(\sigma_0=10^{4}\).
Bottom row: \(\sigma_0=1000\) with (c) \(\xi=0.10\,\mathrm{rad}\), (d) \(\xi=0.25\,\mathrm{rad}\).
White region: no event horizon (\(g>g_{\max}(a)\)).
Color scale: \(\mathcal{I}(a/M,g/M;\sigma_0,\xi)\) (proportional to \(P_{\rm extr}\) up to the constant prefactor \(2\pi w_0U_{\rm in}\)); values corresponding to \(P_{\rm extr}<P_{\min}\) in our fiducial normalization are clipped to zero (dark blue) and are not used to set extraction-based bounds.}
\label{fig:ExtractionCA-Map-2x2}
    \end{figure*}

In the maps below (Fig.~\ref{fig:ExtractionCA-Map-2x2}), we display only the geometric/kinematic proxy \(\mathcal{I}\); the overall multiplicative factor \(2\pi\,w_0\,U_{\rm in}\) required to convert \(\mathcal{I}\) into \(P_{\rm extr}\) is therefore absorbed into the color scale and discussed separately when quoting physical units. Moreover, when producing the extraction maps a practical detectability floor at \(P_{\min}\) is implemented: using a fiducial normalization for $2\pi w_0U_{\rm in}$, we set $P_{\rm extr}\to0$ whenever the implied power satisfies $P_{\rm extr}<P_{\min}$, in order to prevent visually and statistically negligible extraction from driving the subsequent discrepancy-based scans.

For reference, the dimensional normalization in \(P_{\rm extr}=2\pi\,w_0\,U_{\rm in}\,\mathcal{I}\) is intrinsically model-dependent, because it depends on the (poorly constrained) upstream enthalpy density \(w_0\) and inflow speed \(U_{\rm in}\) in the reconnection layer.
At the present stage, our goal is therefore not a sharp prediction for \(P_{\rm extr}\), but an order-of-magnitude anchoring consistent with established jet-power benchmarks. 
A natural reference scale is provided by the Blandford--Znajek (BZ) mechanism and by GRMHD modeling/observational inferences for nearby systems.
For instance, EHT-based modeling and multi-scale constraints for M87* point to jet powers spanning \(\sim10^{42}\)–\(10^{44}\,{\rm erg\,s^{-1}}\) depending on the tracer/scale considered~\cite{EHTVIII2021}.
Motivated by these benchmarks, and consistently with the fact that fast reconnection can be competitive with (or exceed) BZ in an intermediate magnetization range while remaining subdominant at very large \(\sigma_0\)~\cite{ComissoAsenjoPRD2021},
we adopt \(P_{\rm extr}\sim10^{40}\)–\(10^{44}\,{\rm erg\,s^{-1}}\) as an illustrative range for \(M\sim10^{6}\)–\(10^{9}M_\odot\) and high spin.
Importantly, this choice affects only the overall multiplicative normalization: the maps in terms of \(\mathcal{I}\) are unchanged and can be rescaled a posteriori as improved constraints on \(w_0\) and \(U_{\rm in}\) become available.



In addition to the purely geometric frontier $g_{\max}(a)$, we formulate an \emph{extraction-based} constraint on $g/M$ by comparing, at fixed spin $a/M$ and fixed $(\sigma_0,\xi)$, the extracted power in the NED-deformed geometry with the Kerr benchmark. Using Eq.~\eqref{eq:Pextr_equatorial}, we define the relative discrepancy
\begin{equation}
\Delta_P(a;g\mid\sigma_0,\xi)\;\equiv\;
\frac{\big|P_{\rm extr}(a,g;\sigma_0,\xi)-P_{\rm extr}(a,0;\sigma_0,\xi)\big|}
{P_{\rm extr}(a,0;\sigma_0,\xi)}\,,
\label{eq:DeltaP_def}
\end{equation}
and the corresponding $\delta$--dependent upper bound
\begin{equation}
\begin{aligned}
g_\delta(a\mid\sigma_0,\xi)
&\equiv
\sup\Big\{\,g\in[0,g_{\max}(a)):\\
&\hspace{2.2em}\Delta_P(a;g\mid\sigma_0,\xi)\le \delta\,\Big\}.
\end{aligned}
\label{eq:gdelta_def}
\end{equation}
Operationally, for each fixed $a/M$ we scan $g/M\in(0,g_{\max}(a))$ and take $g_\delta$ as the largest value for which $\Delta_P\le\delta$. The threshold \(\delta\) controls the stringency of the constraint and may be interpreted as an (idealized) fractional uncertainty on the inferred extracted power, assuming the Kerr prediction of the Comisso--Asenjo model provides the calibration baseline. In what follows we adopt \(\delta=0.3\) as a representative choice of tolerance, stressing that this is not a claim on current observational accuracy with respect to the Kerr case ($g=0$). 
To keep the extraction-based comparison focused on the spacetime sector, we use representative, radially constant reference values for the plasma inputs (in particular \(w_0\) and \(U_{\rm in}\)). In the ratio~\eqref{eq:DeltaP_def} the overall factor \(2\pi\) cancels identically, and any residual dependence on \(w_0U_{\rm in}\) cancels as well provided these quantities vary only weakly with \(g\) at fixed \(a/M\). Under this working hypothesis, the bound \(g_\delta(a)\) is controlled primarily by the geometric/kinematic contributions entering \(P_{\rm extr}\), namely the redshift and frame dragging encoded in \(\epsilon_-^\infty(r)\), the proper-area measure \(\sqrt{\gamma_{rr}\gamma_{\phi\phi}}\), and the \(g\)-induced shifts of \((r_H,r_L)\) together with the extraction mask \(f_{\rm extr}(r)\).

To quote powers in cgs units, we reinstate the dimensional prefactor in Eq.~\eqref{eq:Pextr_equatorial} and write \(P_{\rm phys}=2\pi w_0U_{\rm in}\mathcal{I}\). In practice, for each \(a/M\) we report \(P_{\rm phys}\) (in units of \(10^{40}\,{\rm erg\,s^{-1}}\)) for Kerr (\(g=0\)) and for representative NED values chosen near (or below) \(g_\delta(a)\), using the fiducial ranges for \(w_0\), \(B_0\), and \(U_{\rm in}\) discussed above. Since the dependence on \(w_0U_{\rm in}\) is purely multiplicative, all plotted amplitudes scale linearly with this product, while the \((a/M,g/M)\), \(\sigma_0\), and \(\xi\) dependence is entirely encoded in \(\mathcal{I}\).
\vspace{10pt}


The next figure (Fig.~\ref{fig:gdelta_3x2}) shows the extraction--based constraint as curves \(g_\delta(a)/M\) versus \(a/M\), obtained from the tolerance condition in Eqs.~\eqref{eq:DeltaP_def}--\eqref{eq:gdelta_def}. For reference we also plot the relaxed bound \(g_{4\delta}(a\mid\sigma_0,\xi)\), defined by \(\delta\to4\delta\), which brackets how the tolerated discrepancy propagates into the allowed range of \(g/M\). The left panel fixes the field orientation at \(\xi=0\) and compares different values of the magnetization \(\sigma_0\), while the right panel fixes \(\sigma_0=5000\) and compares three nearly azimuthal launch angles, \(\xi=0\), \(0.10\), and \(0.25\). In both panels, the geometric ceiling associated with horizon existence is also shown for reference, together with the corresponding black--hole domain. For a given spin, the admissible values of \(g/M\) lie below the corresponding \(g_\delta(a)/M\) curve and, necessarily, below the geometric ceiling. Overall, the bound tightens toward high spin—where frame dragging and the ergoregion maximize the negative--energy window—whereas at moderate spin it becomes more sensitive to \((\sigma_0,\xi)\): modest increases in \(\xi\) can quench extraction, and for \(\sigma_0\lesssim\mathcal{O}(10^2\text{--}10^3)\) the window can close already around \(a/M\sim0.75\) in our grids, leaving only weak or null bounds except near extremal rotation. The sensitivity to the angle \(\xi\) can also be attributed to the simplicity of the model assumed here: our single--layer equatorial construction is intentionally selective and thus emphasizes the dependence on the azimuthal projection \(\cos\xi\). Presumably, in a more realistic 3D configuration, a distribution of reconnection sites and launch angles would broaden the effective angular acceptance while preserving the underlying geometric sensitivity.


\begin{figure*}[t]
    \centering
    \subfigure{
        \includegraphics[height=5.5cm]{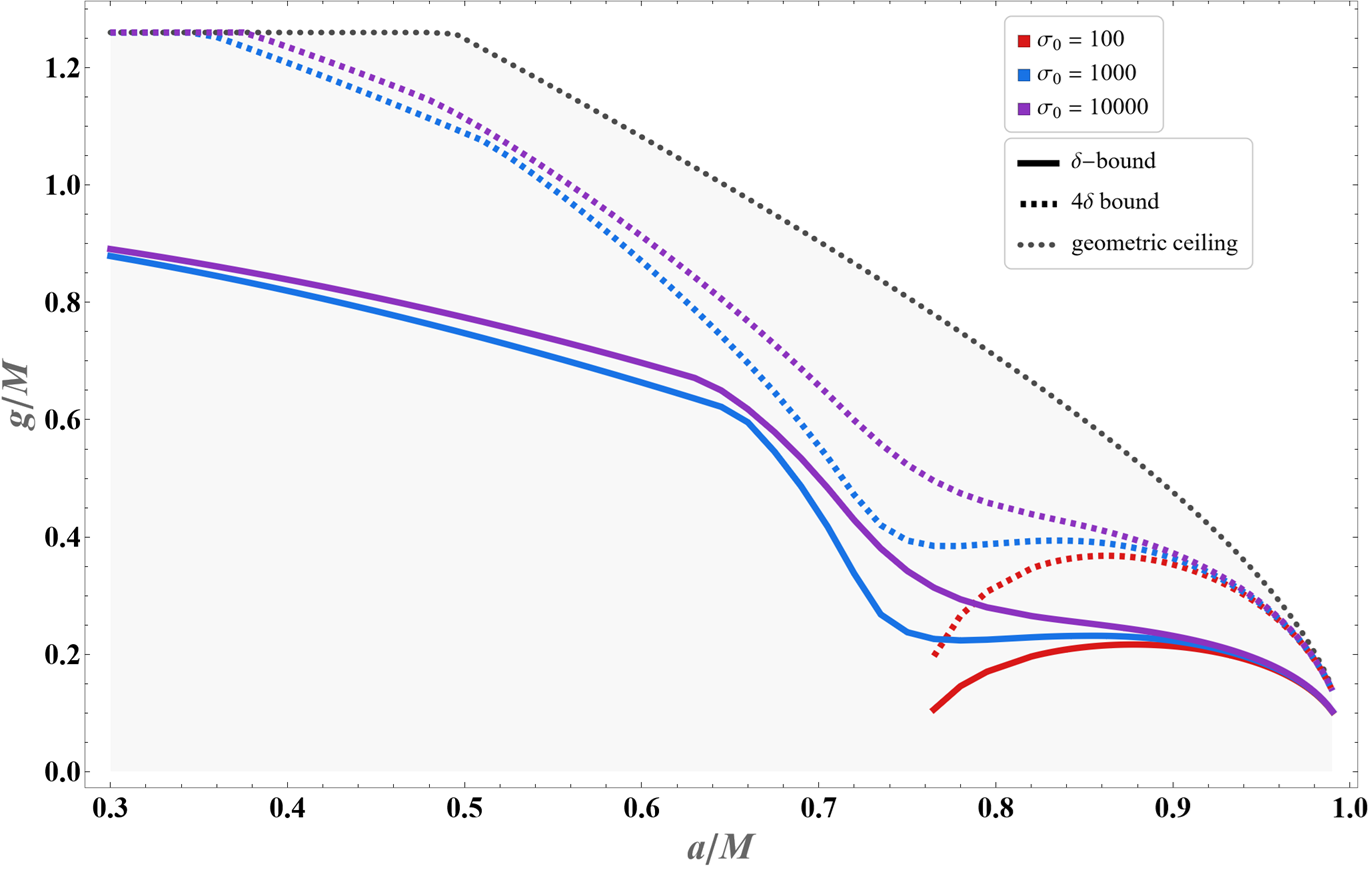}
    }\hfill
    \subfigure{
        \includegraphics[height=5.5cm]{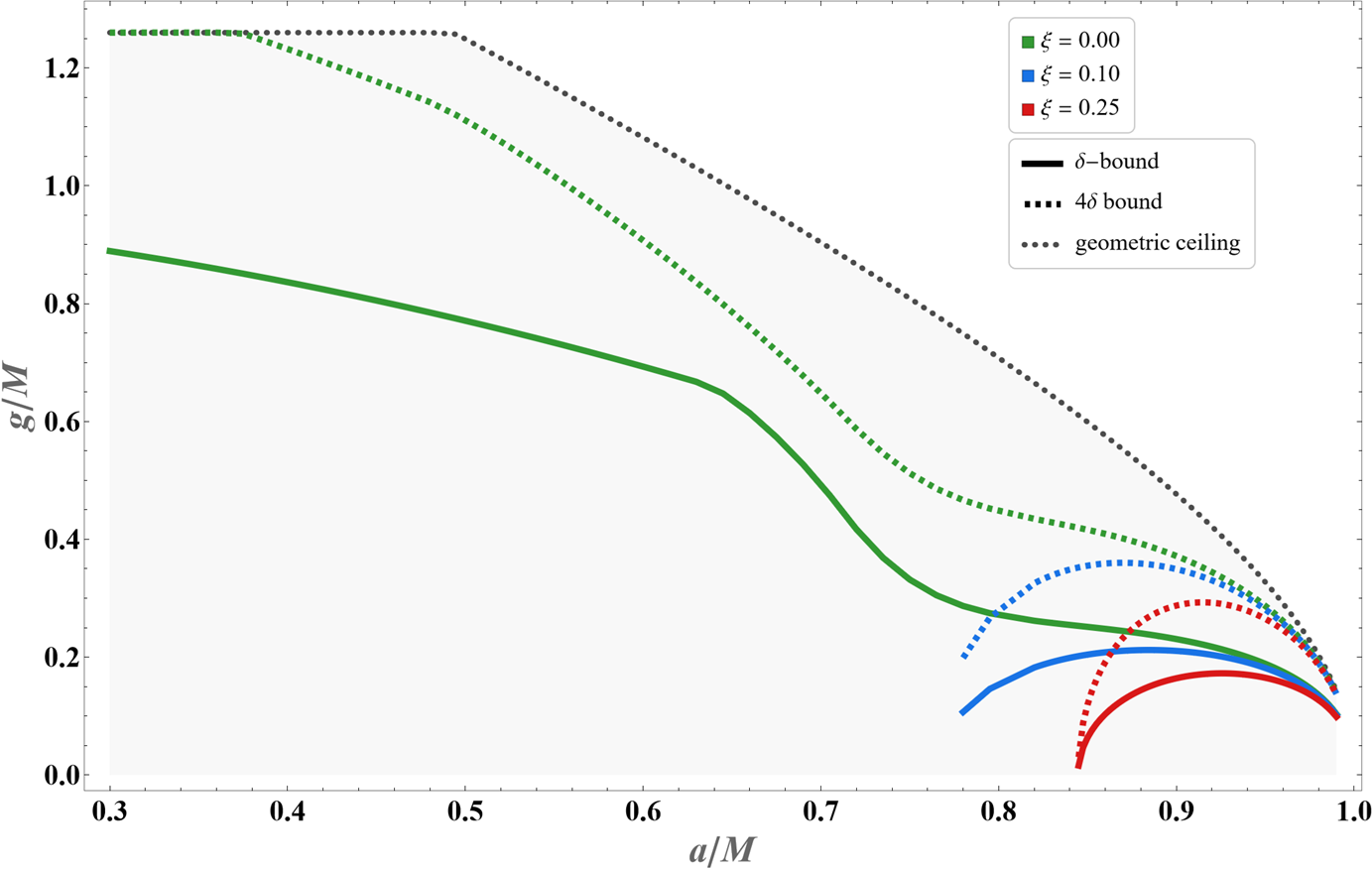}
    }
    \caption{\small\textbf{Extraction--based bounds on \(g/M\).}
The panels show the upper bound \(g_\delta(a\mid\sigma_0,\xi)/M\) defined by Eq.~\eqref{eq:gdelta_def}, inferred from the power-discrepancy criterion in Eqs.~\eqref{eq:DeltaP_def}--\eqref{eq:gdelta_def}. Solid curves denote the representative tolerance \(\delta=0.3\), while dashed curves show the relaxed choice \(4\delta\). The dotted curve marks the geometric ceiling associated with horizon existence, and the lightly shaded region below it indicates the black--hole domain. Left panel: dependence on the magnetization \(\sigma_0\) at fixed field orientation \(\xi=0\), with \(\sigma_0=100,1000,10000\). Right panel: dependence on the orientation \(\xi\) at fixed magnetization \(\sigma_0=5000\), with \(\xi=0,0.10,0.25\).}
\label{fig:gdelta_3x2}
    \end{figure*}

A few caveats about measurability are nonetheless essential. 
The constrained quantity is the reconnection--driven contribution to the near--horizon energy budget, whereas the observed jet/outflow power generally includes additional channels, notably Blandford--Znajek extraction and accretion--powered components. Separating these contributions requires modeling and introduces degeneracies with the plasma normalization \(w_0\), the magnetization \(\sigma_0\), the outflow geometry \(\xi\), and the magnetic--flux state (MAD vs SANE)~\cite{McKinney2004,Komissarov2005,Krolik2012}.
In practice, achieving sub--\(10\%\) accuracy on the reconnection fraction is not currently realistic: uncertainties in the horizon--threading magnetic flux \(\Phi_{\rm BH}\), radiative efficiency and particle content, Doppler beaming and inclination, and the accretion state typically yield fractional errors \(\gtrsim 30\%\) even in the best--characterized sources (and of order unity otherwise). For this reason we formulate constraints through a tolerated fractional discrepancy \(\delta\) in \(\Delta_P\), which maps to \(g_\delta(a)\) without relying on a precisely known absolute normalization. In favorable high--spin MAD systems with independent spin and \(\Phi_{\rm BH}\) estimates and dense polarimetric/timing coverage, next--generation data may plausibly push to \(\delta\lesssim 0.1\), but a full joint--inference implementation is beyond the scope of this geometry--focused study.

Two practical points qualify the interpretation of our scans. 
First, we introduce a conservative detectability threshold \(P_{\min}\) (here \(3\times10^{38}\,\mathrm{erg\,s^{-1}}\)) and implement it as a clipping in the maps: whenever \(P_{\rm extr}<P_{\min}\) we set \(P_{\rm extr}\to0\) (dark-blue pixels). This threshold is used only to prevent extraction-based bounds from being driven by numerically non-robust or observationally imperceptible power levels; it does not modify the geometric domain \(g<g_{\max}(a)\).
Second, the ratio \(\Delta_P\) in Eq.~\eqref{eq:DeltaP_def} mitigates uncertainties in the prefactors provided that they are only weakly \(g\)--dependent at fixed \(a/M\).
Finally, near--horizon reconnection timescales \(t_{\rm rec}\!\sim\!L/(\mathcal{R}v_A)\) (hours for SMBHs and milliseconds for stellar--mass BHs at our fiducial \(L\)) suggest that quasi--steady segments of activity can be isolated in long monitoring campaigns, making high--spin sources with stable accretion states the most promising targets. A tighter, population--level program would naturally combine multi--epoch power proxies with polarization/variability diagnostics that tag reconnection episodes; a systematic treatment of such joint inference is deferred to future work.

\subsection{Blandford--Znajek benchmark and geometric imprint of the NED parameter}
\label{subsec:BZ_benchmark}

Before concluding, it is useful to place the Comisso--Asenjo (CA) reconnection channel in context by comparing it with the Blandford--Znajek (BZ) mechanism, where black-hole rotational energy is extracted electromagnetically through a magnetic field threading the event horizon. This comparison is standard in the literature (see e.g.\ \cite{ComissoAsenjoPRD2021,CarleoEPJC2022}) and it is particularly informative in our setting: CA relies on localized dissipation inside the ergoregion, whereas BZ is a large-scale, horizon-anchored electromagnetic torque, yet both channels inherit a genuine dependence on the background geometry through redshift, frame dragging, and the horizon location.

For the CA channel we have already introduced the enthalpy-normalized proxy $\mathcal{I}(a/M,g/M;\sigma_0,\xi)$ in Eq.~\eqref{eq:proxy_I}, with all $(a/M,g/M)$ dependence carried by the integral over the ergoregion through $\epsilon_-^\infty(r)$, the proper-area measure, the $g$-dependent endpoints $(r_H,r_L)$, and the extraction mask $f_{\rm extr}(r)$.
In the following we fix a representative field orientation and focus on the dependence on $(a/M,g/M)$ and $\sigma_0$; specifically we set $\xi=0$ (nearly azimuthal launch), which maximizes the coupling to frame dragging through $\cos\xi$ in Eq.~\eqref{eq:epsiloninfty} and provides a clean benchmark.

For the BZ channel, we adopt the standard high-spin generalization of the original Blandford--Znajek scaling, widely used in membrane-paradigm/force-free treatments and GRMHD-motivated fits \cite{BlandfordZnajek1977,ThornePriceMacdonald1986,TchekhovskoyNarayanMcKinney2010,Tchekhovskoy2011,ComissoAsenjoPRD2021,CarleoEPJC2022}:
\begin{equation}
P_{\rm BZ}(a,g)\;\simeq\;\kappa_{\rm BZ}\,\Phi_{\rm BH}^2\,\Omega_H^2
\Big[1+\chi\,(\Omega_H M)^2+\zeta\,(\Omega_H M)^4\Big]\,,
\label{eq:PBZ_poly}
\end{equation}
where $\Phi_{\rm BH}$ is the magnetic flux threading one hemisphere of the horizon and $\Omega_H$ is the horizon angular velocity. We work in units $G=c=1$, so that $\Omega_H$ has dimensions $1/M$ and $\Omega_H M$ is dimensionless. We adopt the GRMHD-calibrated high-spin fit of Tchekhovskoy, Narayan \& McKinney \cite{TchekhovskoyNarayanMcKinney2010}, i.e.\ $\chi=1.38$ and $\zeta=-9.2$, while $\kappa_{\rm BZ}$ encodes the large-scale magnetospheric geometry. Following \cite{ComissoAsenjoPRD2021,CarleoEPJC2022}, we define the hemisphere flux as
\begin{equation}
\Phi_{\rm BH}\;\equiv\;\frac{1}{2}\int_{\rm hem}\! |B^r|\;dA_{\theta\phi}\,,
\label{eq:PhiBH_def}
\end{equation}
with $dA_{\theta\phi}=\sqrt{g_{\theta\theta}g_{\phi\phi}-g_{\theta\phi}^2}\,d\theta\,d\phi
\quad \text{at }\; r=r_H$.

The dimensionless constants $\kappa_{\rm BZ},\chi,\zeta$ encode the large-scale magnetospheric geometry and the high-spin correction. In particular, $\chi=1.38$ and $\zeta=-9.2$ correspond to the GRMHD-calibrated high-spin expansion reported by Tchekhovskoy, Narayan \& McKinney~\cite{TchekhovskoyNarayanMcKinney2010} (and subsequently adopted in comparative CA/BZ analyses, e.g.\ \cite{ComissoAsenjoPRD2021,CarleoEPJC2022}); the overall factor $\kappa_{\rm BZ}$ is typically at the few $\times 10^{-2}$ level depending on the field geometry.

Crucially for our purposes, the NED parameter $g/M$ enters the BZ channel already at the kinematic level through $\Omega_H(a,g)$ and the horizon location $r_H(a,g)$.
In our notation, the horizon angular velocity is simply the ZAMO frame-dragging frequency evaluated at the horizon,
\begin{equation}
\Omega_H(a,g)=\omega(r_H;a,g),
\label{eq:OmegaH_as_omega}
\end{equation}
with $\omega$ defined in Eq.~\eqref{eq:ZAMOdefs} and $r_H(a,g)$ determined by the horizon condition $\Delta(r_H)=0$ in the rotating NED geometry.
Even if the horizon-threading flux were kept fixed, the BZ luminosity inherits a genuine $g/M$ dependence already at the kinematic level, through the $g$-dependence of the horizon location $r_H(a,g)$ and of the horizon angular velocity $\Omega_H(a,g)$ entering Eq.~\eqref{eq:PBZ_poly}.

To isolate this geometric imprint in a minimally model-dependent way, we follow the same philosophy adopted in comparative CA/BZ studies \cite{ComissoAsenjoPRD2021,CarleoEPJC2022} and treat the horizon-threading flux $\Phi_{\rm BH}$ as an external control parameter, i.e.\ we do not prescribe its response under deformations of the background geometry when varying $g/M$ at fixed $a/M$.
Using Eqs.~\eqref{eq:proxy_I} and \eqref{eq:PBZ_poly}, the CA/BZ comparison can be written as
\[
\frac{P_{\rm extr}(a,g;\sigma_0,\xi=0)}{P_{\rm BZ}(a,g)}
=
\frac{2\pi\,w_0\,U_{\rm in}}{\kappa_{\rm BZ}\,\Phi_{\rm BH}^2}\;
\mathcal{G}(a/M,g/M;\sigma_0),
\]
where the reduced function $\mathcal{G}$ collects the geometry-driven dependence on $(a/M,g/M)$,
\begin{equation}
\begin{aligned}
&\mathcal{G}(a/M,g/M;\sigma_0)\;\equiv\;\\
&\;\;\,\frac{\mathcal{I}(a/M,g/M;\sigma_0,\xi=0)}
{\Omega_H(a,g)^2\Big[1+\chi\,(\Omega_H(a,g) M)^2+\zeta\,(\Omega_H(a,g) M)^4\Big]}\,.
\end{aligned}
\label{eq:G_reduced}
\end{equation}

By construction, $\mathcal{G}$ is the reduced, geometry-driven indicator entering the CA/BZ ratio once the plasma normalization and the large-scale magnetospheric prefactors are collected into the single multiplicative coefficient $(2\pi w_0U_{\rm in})/(\kappa_{\rm BZ}\Phi_{\rm BH}^2)$.

A residual modeling freedom concerns how the horizon-threading flux $\Phi_{\rm BH}$ responds when scanning the upstream magnetization $\sigma_0$.
We therefore consider two complementary prescriptions.
In a first, minimally committal choice, $\Phi_{\rm BH}$ is treated as an external control parameter and kept fixed across the scan, so that the $g/M$ dependence of $\mathcal{G}$ arises purely from spacetime kinematics (through $\Omega_H$) and from the ergoregion integral defining $\mathcal{I}$.

\begin{figure*}[t]
    \centering
    \subfigure{
        \includegraphics[height=5cm]{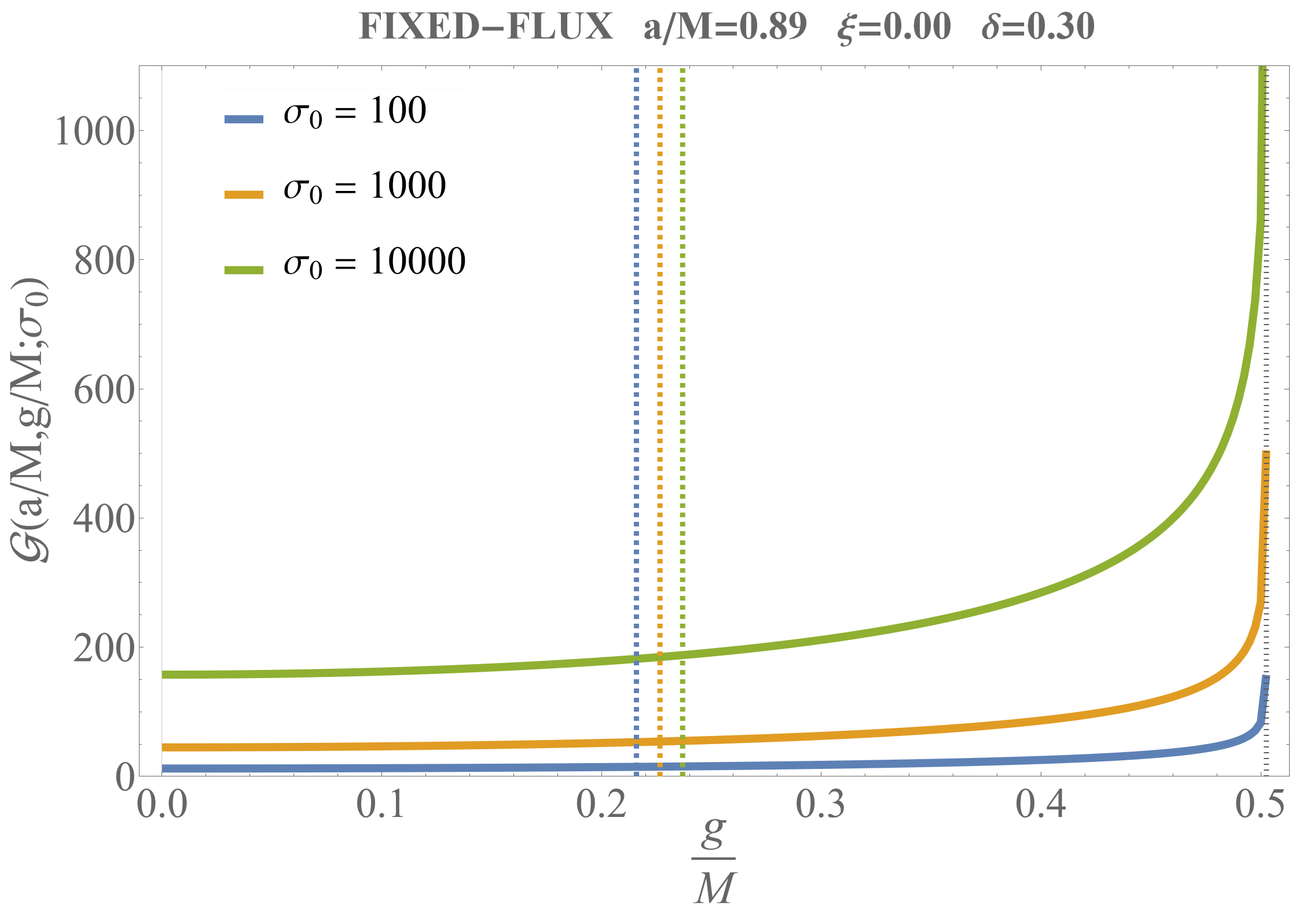}
    }
     \subfigure{
        \includegraphics[height=5cm]{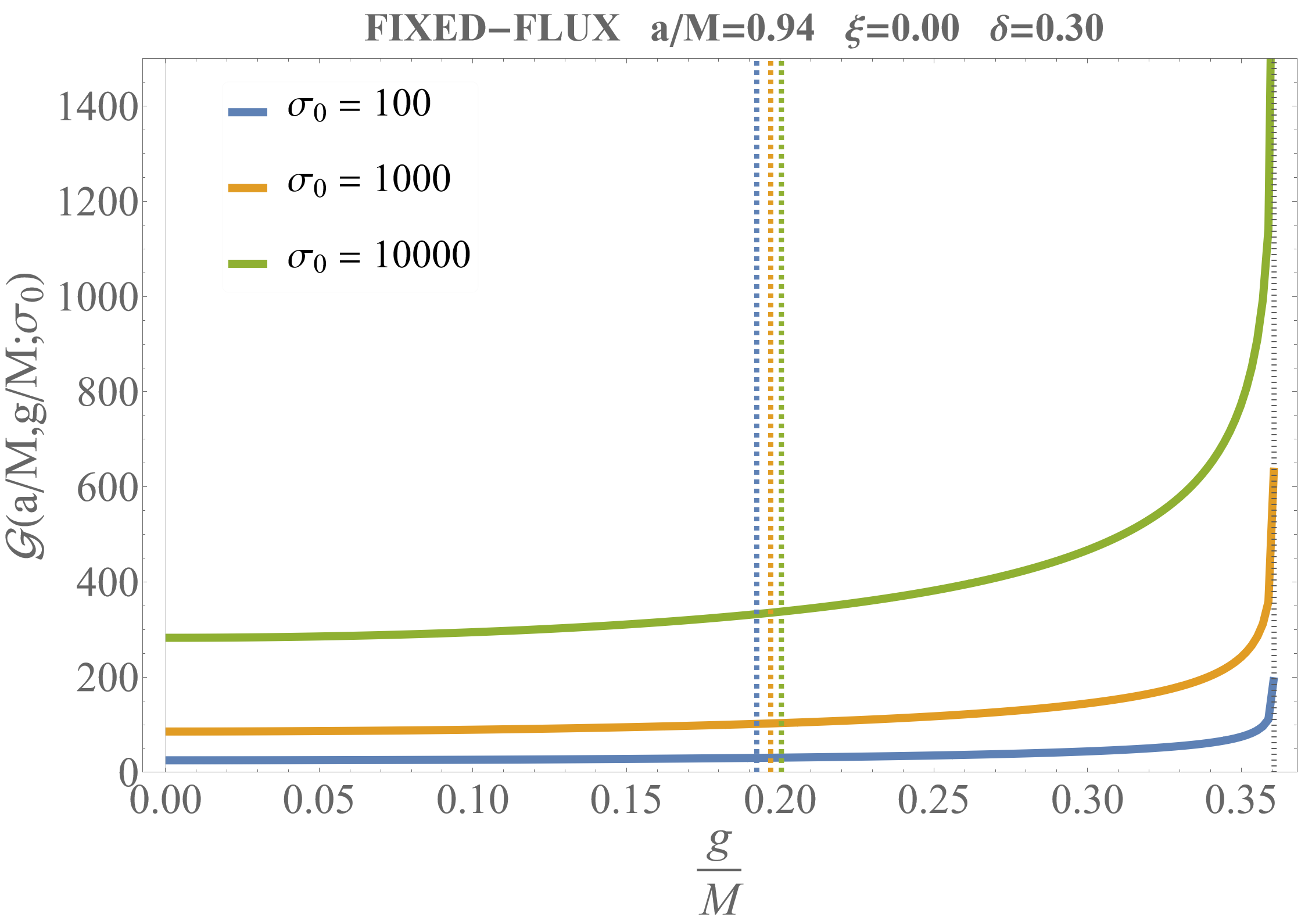}
    }
    \subfigure{
        \includegraphics[height=5cm]{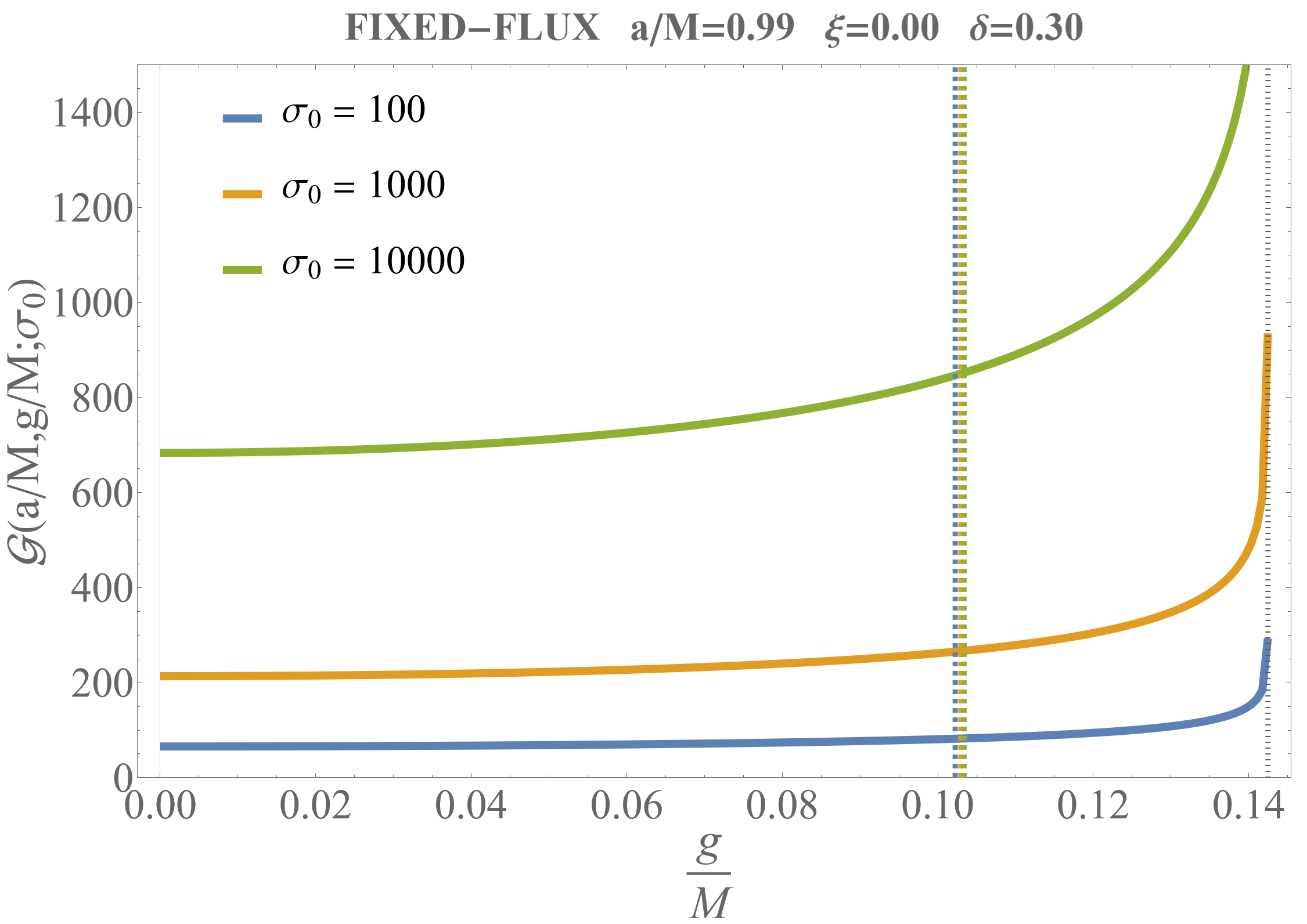}
    }
    \subfigure{
        \includegraphics[height=5cm]{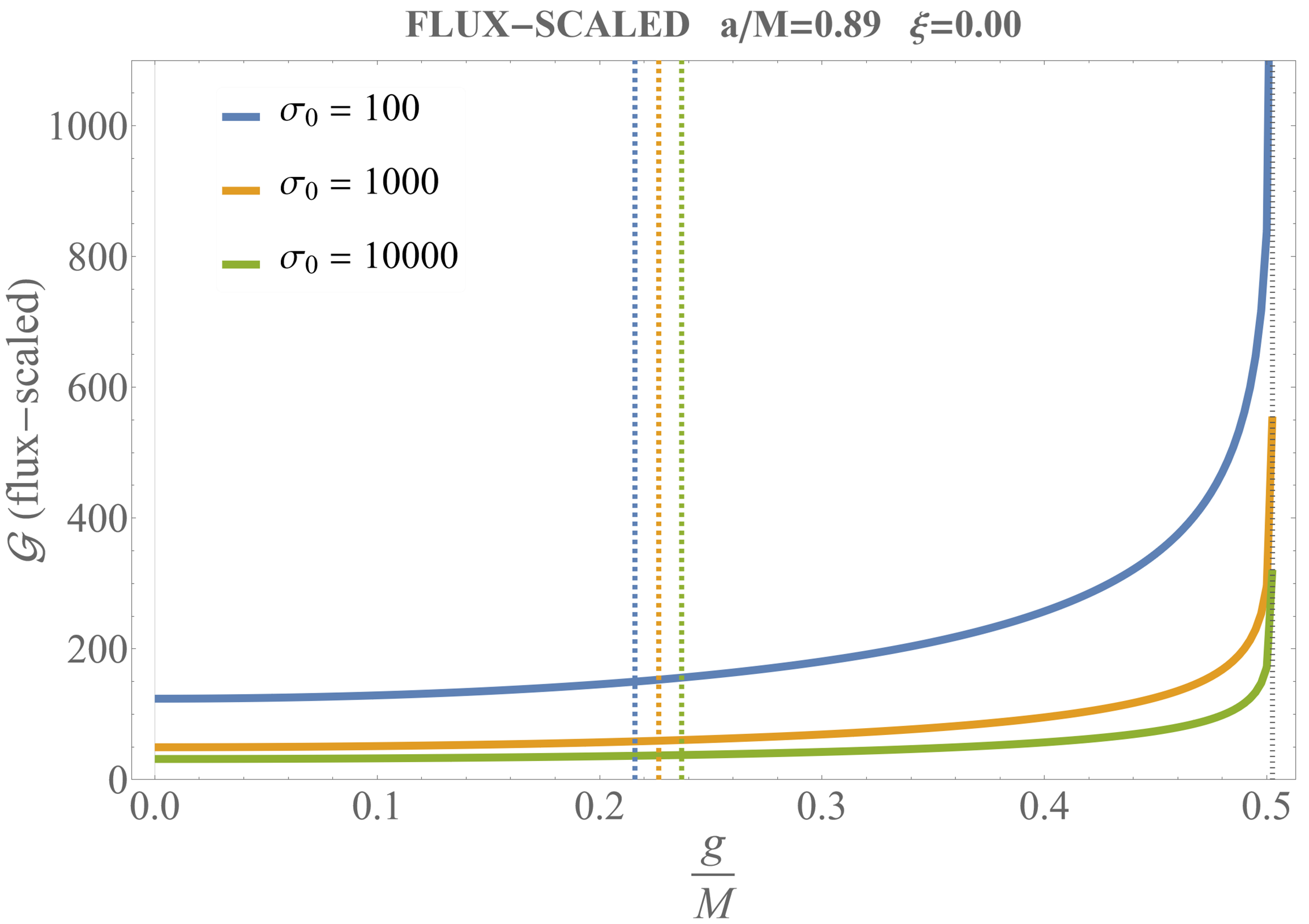}
    }
    \subfigure{
        \includegraphics[height=5cm]{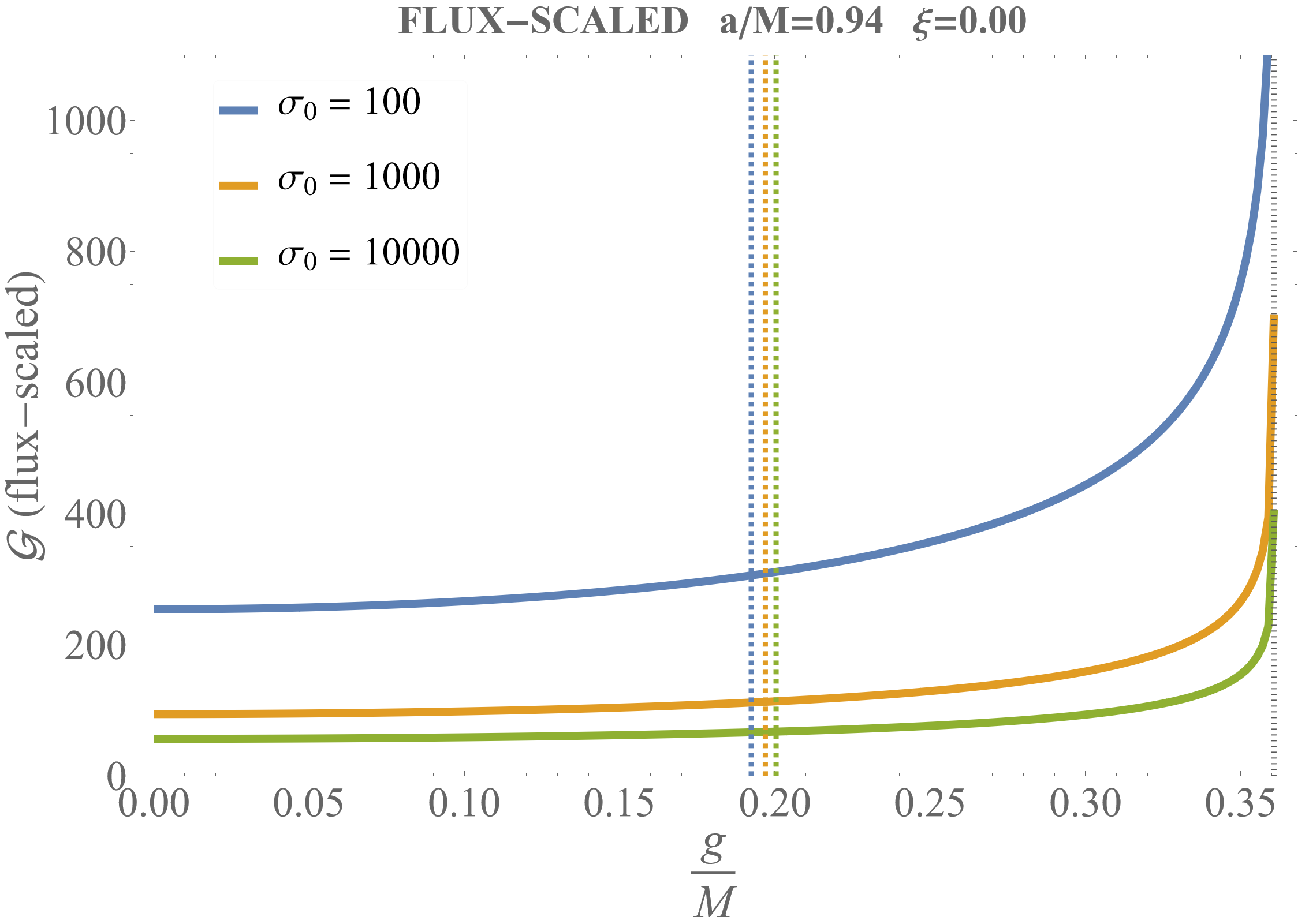}
    }
    \subfigure{
        \includegraphics[height=5cm]{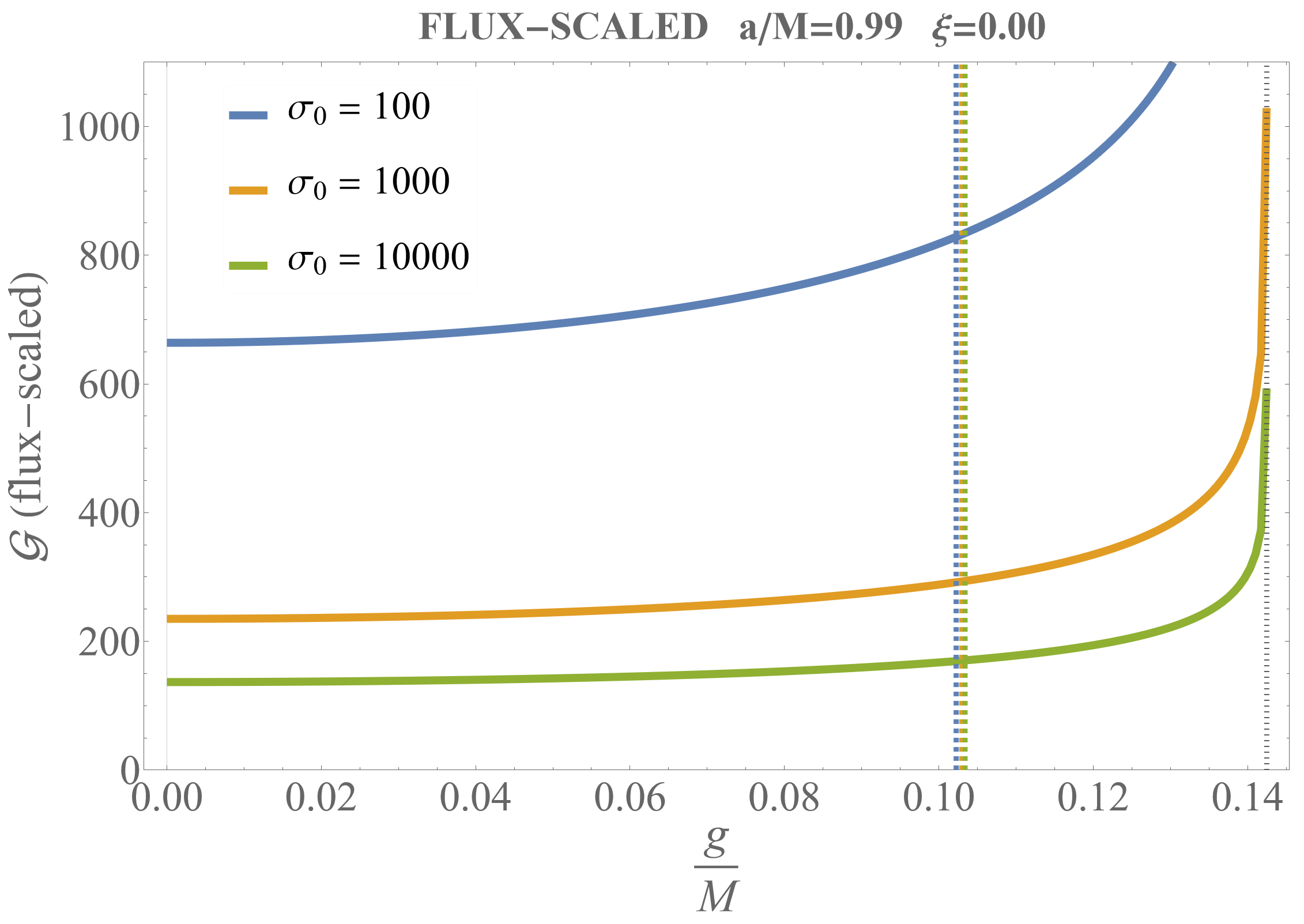}
    }
    \caption{\small
Reduced CA/BZ comparison versus $g/M$, shown through the geometry-driven ratio
$\mathcal{G}(a/M,g/M;\sigma_0)=\mathcal{I}(a/M,g/M;\sigma_0,\xi=0)\Big/\!\left(\Omega_H^2\left[1+\chi(\Omega_H M)^2+\zeta(\Omega_H M)^4\right]\right)$
for $\sigma_0=10^2,10^3,10^4$ (solid curves).
Panels (a)--(c) correspond to the fixed-flux benchmark, i.e.\ $\Phi_{\rm BH}$ is treated as an external control parameter, so that the $g/M$ dependence arises solely from the spacetime kinematics entering $\mathcal{I}$ and $\Omega_H$.
Panels (d)--(f) illustrate a minimal flux-response ansatz (meant to probe qualitative trends) in which the BZ normalization is rescaled as
$\Phi_{\rm BH}^2\to \Phi_{\rm BH}^2\,\tilde{\Phi}^2(\sigma_0)$ with
$\tilde{\Phi}^2(\sigma_0)=(\sigma_0/\sigma_{\rm ref})/[1+\sigma_0/\sigma_{\rm sat}]$; accordingly, the plotted quantity is $\mathcal{G}/\tilde{\Phi}^2(\sigma_0)$.
We fix $\sigma_{\rm ref}=10^3$ and $\sigma_{\rm sat}=10^4$.
In each panel, colored dashed vertical lines mark the maximal $g/M$ allowed by the extractive bound
$\Delta_P(a;g\mid\sigma_0,\xi)\le\delta$ with $\delta=30\%$ (one bound per $\sigma_0$),
while the grey dotted line marks the geometric ceiling set by horizon existence.}
\label{fig:Gred_2x3_panel}
    \end{figure*}

In a second, deliberately schematic variant meant to probe trends (rather than to provide a GRMHD closure), we allow the horizon-threading flux to increase with magnetization while enforcing saturation at high $\sigma_0$, as suggested by the flux-limited behavior of magnetically arrested (MAD) configurations. Concretely, we rescale the BZ normalization as $\Phi_{\rm BH}^2\!\to\!\Phi_{\rm BH}^2\,\tilde{\Phi}^2(\sigma_0)$, with
\begin{equation}
\tilde{\Phi}^2(\sigma_0)
=\frac{\sigma_0/\sigma_{\rm ref}}{1+\sigma_0/\sigma_{\rm sat}}\,,
\qquad
\sigma_{\rm ref}=10^3,
\label{eq:Phi_response_sat}
\end{equation}
so that $\tilde{\Phi}^2\simeq\sigma_0/\sigma_{\rm ref}$ for $\sigma_0\ll\sigma_{\rm sat}$ (consistent with $\sigma_0=b_0^2/w_0$ at fixed $w_0$), while $\tilde{\Phi}^2$ saturates to $\sigma_{\rm sat}/\sigma_{\rm ref}$ for $\sigma_0\gg\sigma_{\rm sat}$.
Here $\sigma_{\rm ref}$ is a purely conventional reference value (it only fixes the overall normalization of the flux-scaled benchmark), whereas $\sigma_{\rm sat}$ controls the onset of flux saturation, mimicking the regulation of the accumulated horizon flux by interchange/eruption dynamics in high-flux states. Unless stated otherwise, we set $\sigma_{\rm sat}=10^4$ in Fig.~\ref{fig:Gred_2x3_panel}.

Figure~\ref{fig:Gred_2x3_panel} shows that $\mathcal{G}(a/M,g/M;\sigma_0)$ remains relatively flat at small $g/M$, while exhibiting a mild but systematic upward drift as $g/M$ increases across the explored domain (within the extraction-based window $g\le g_\delta(a\mid\sigma_0,\xi)$). In the fixed-flux normalization [panels (a)--(c)], this indicates that the NED deformation enhances the CA proxy $\mathcal{I}$ slightly more than it enhances the purely kinematic BZ factor encoded in $\Omega_H(a,g)$, so that the reduced CA/BZ comparison is mildly favored at larger $g/M$. 

Panels (d)--(f) provide a complementary viewpoint: once a monotonic increase of the horizon-threading flux with magnetization is allowed (even in this minimalist form), the BZ benchmark steepens with $\sigma_0$ and the curve ordering correspondingly inverts, making the BZ channel dominant at large $\sigma_0$, consistently with the quadratic dependence on $\Phi_{\rm BH}$ in Eq.~\eqref{eq:PBZ_poly} and with the qualitative CA/BZ trends reported in~\cite{ComissoAsenjoPRD2021,CarleoEPJC2022}. Taken together, the two normalizations support a consistent qualitative message: the $g/M$ dependence is modest but coherent in sign, whereas the absolute hierarchy between CA and BZ is governed primarily by the large-scale magnetic-flux state (encoded in $\kappa_{\rm BZ}$ and $\Phi_{\rm BH}$), rather than by the spacetime deformation alone.

Therefore, while the Blandford--Znajek process provides a natural benchmark for horizon-anchored electromagnetic extraction, the Comisso--Asenjo mechanism offers an ergoregion-localized, geometry-sensitive probe.
The extraction-based bound $g_\delta(a)$ derived from the CA channel is thus complementary in the sense of probing a different geometric weighting of the same background deformation, and it can be combined with other observables within a multi-channel assessment of Kerr-like departures.

\section{\label{sec:level5} Conclusions and outlook}

This work advances a two–track assessment of rotating black holes in general relativity coupled to nonlinear electrodynamics (NED). On the geometric side, starting from Eqs.~\eqref{eq:NEDmetric}--\eqref{eq:NEDmetric3} we derived closed–form ZAMO scalars, equatorial kinematics, and the horizon/ergoregion structure as functions of $(a/M,g/M)$, computed the corresponding quasi–normal modes (QNMs), and defined a ringdown ceiling $g_{\rm QNM}(a)$ that is consistent, in the eikonal limit, with the light–ring/shadow ceiling $g_{\rm sh}(a)$. On the dynamical side, we cast the Comisso–Asenjo (CA) reconnection channel in ZAMO variables and, from the energy at infinity per upstream enthalpy in Eq.~\eqref{eq:epsiloninfty}, delineated the extraction window and integrated the equatorial power in Eq.~\eqref{eq:Pextr_equatorial}, introducing the extraction–based bound $g_\delta(a\mid\sigma_0,\xi)$ via the discrepancy measure $\Delta_P$ in Eq.~\eqref{eq:DeltaP_def} without assuming monotonicity of $\Delta_P(a;g)$.

To place this scale into context and to isolate purely geometric effects, we compared CA with the Blandford–Znajek (BZ) benchmark through the reduced indicator $\mathcal{G}(a/M,g/M;\sigma_0)\equiv
\mathcal{I}\Big/\!\left(\Omega_H^2\left[1+\chi(\Omega_H M)^2+\zeta(\Omega_H M)^4\right]\right)$ defined in Eq.~\eqref{eq:G_reduced}. Since $g/M$ enters the BZ luminosity kinematically via $r_H(a,g)$ and $\Omega_H(a,g)$ [Eq.~\eqref{eq:PBZ_poly}], $\mathcal{G}$ exhibits a modest but coherent $g/M$ dependence across both the fixed–flux case and a minimal flux–response variant, while the absolute CA/BZ hierarchy is governed primarily by the horizon–threading flux $\Phi_{\rm BH}$ rather than by the spacetime deformation alone.

The two routes are therefore complementary: QNMs/shadow provide robust ceilings when reconnection is disfavored (low $\sigma_0$ or non–azimuthal exhausts), whereas in favorable plasma states (high magnetization and nearly azimuthal launch) the extraction–based $g_\delta(a\mid\sigma_0,\xi)$ becomes comparably stringent—or tighter at high spin—probing the deformation \emph{inside} the ergoregion and independently of the horizon–anchored torque. Throughout, we state and use working assumptions transparently (single equatorial layer, Maxwell limit in the diffusion region, test–fluid on a fixed background, representative tolerance $\delta$) and employ ratios such as $\Delta_P$ to cancel uncertain normalizations.

From an observational perspective, this framework yields testable inequalities once three ingredients are constrained or bracketed: the spin $a/M$, a proxy for the horizon–threading flux $\Phi_{\rm BH}$ (e.g., MAD vs SANE), and broad priors on $(\sigma_0,\xi)$ from GRMHD–informed modeling. For high–spin sources with VLBI–scale polarization and variability, joint fits of multi–epoch power proxies with reconnection diagnostics (polarization swings, rapid flaring) translate into a tolerated discrepancy $\delta$ in $\Delta_P$ and thus into $g_\delta(a\mid\sigma_0,\xi)$; the CA/BZ comparison via the reduced indicator $\mathcal{G}(a/M,g/M;\sigma_0)$ provides an internal consistency check. While current systematics in $\Phi_{\rm BH}$, composition, and radiative efficiency dominate the error budget, forthcoming VLBI and polarimetric data can realistically shrink $\delta$ for a handful of stable, high–spin AGN, enabling population–level bounds that are complementary to shadow/QNM ceilings.

Looking ahead, several avenues naturally present themselves: a near–horizon sufficient criterion for extraction would bracket $g_\delta(a)$ with a closed estimate $g_{\max}^{\rm (extr)}(a/M,\sigma_0)$; embedding the present diagnostics in global GRMHD and kinetic simulations can supply realistic distributions of $(\sigma_0,\xi,U_{\rm in})$ and test the behavior of $\Delta_P(g)$; observationally, combining power proxies with polarization/timing indicators of reconnection and multi–epoch horizon–scale data, and integrating independent inputs (e.g., X–ray reflection spectroscopy or PTA–based priors \cite{Ingram2019}), can enable population–level constraints. Taken together, the geometry–only (QNM/shadow) and dynamics–based (CA, in dialogue with BZ) pathways deliver a coherent, spin–resolved protocol to bound NED deformations of rotating black–hole geometries while keeping astrophysical systematics explicit and under control.

\section*{Acknowledgments}
M. F. thanks Tanmay Kumar Poddar for useful comments and discussions that contributed to improving the manuscript.

G. L., A.~\"Ovg\"un, and R.~C.~Pantig acknowledge networking support from COST Action CA21106 -- COSMIC WISPers in the Dark Universe: Theory, astrophysics and experiments (CosmicWISPers), COST Action CA22113 -- Fundamental challenges in theoretical physics (THEORY-CHALLENGES), COST Action CA21136 -- Addressing observational tensions in cosmology with systematics and fundamental physics (CosmoVerse), COST Action CA23130 -- Bridging high and low energies in search of quantum gravity (BridgeQG), and COST Action CA23115 -- Relativistic Quantum Information (RQI), funded by COST (European Cooperation in Science and Technology).

A.~\"Ovg\"un also acknowledges support from EMU, TUBITAK, ULAKBIM (Turkiye), and SCOAP3 (Switzerland). G. L. acknowledges the Istituto Nazionale di Alta Matematica (INdAM), Gruppo Nazionale di Fisica Matematica.

\appendix

\section{Supplementary geometric diagnostics for the rotating NED black hole}
\label{app:sec3_supplement}

For completeness, we collect here several geometric diagnostics of the rotating NED spacetime introduced in Sec.~\ref{sec:level3}, which were included in an earlier, more extended version of that section. Although they do not enter explicitly into the extraction formalism of Sec.~\ref{sec:level4}, they provide useful intuition on how the deformation parameter \(g/M\) modifies the horizon structure, the ergoregion, and the main equatorial geodesic properties of the rotating background.

\medskip

\begin{figure}[!htbp]
  \centering
  \includegraphics[width=\columnwidth,height=5.5cm,keepaspectratio]{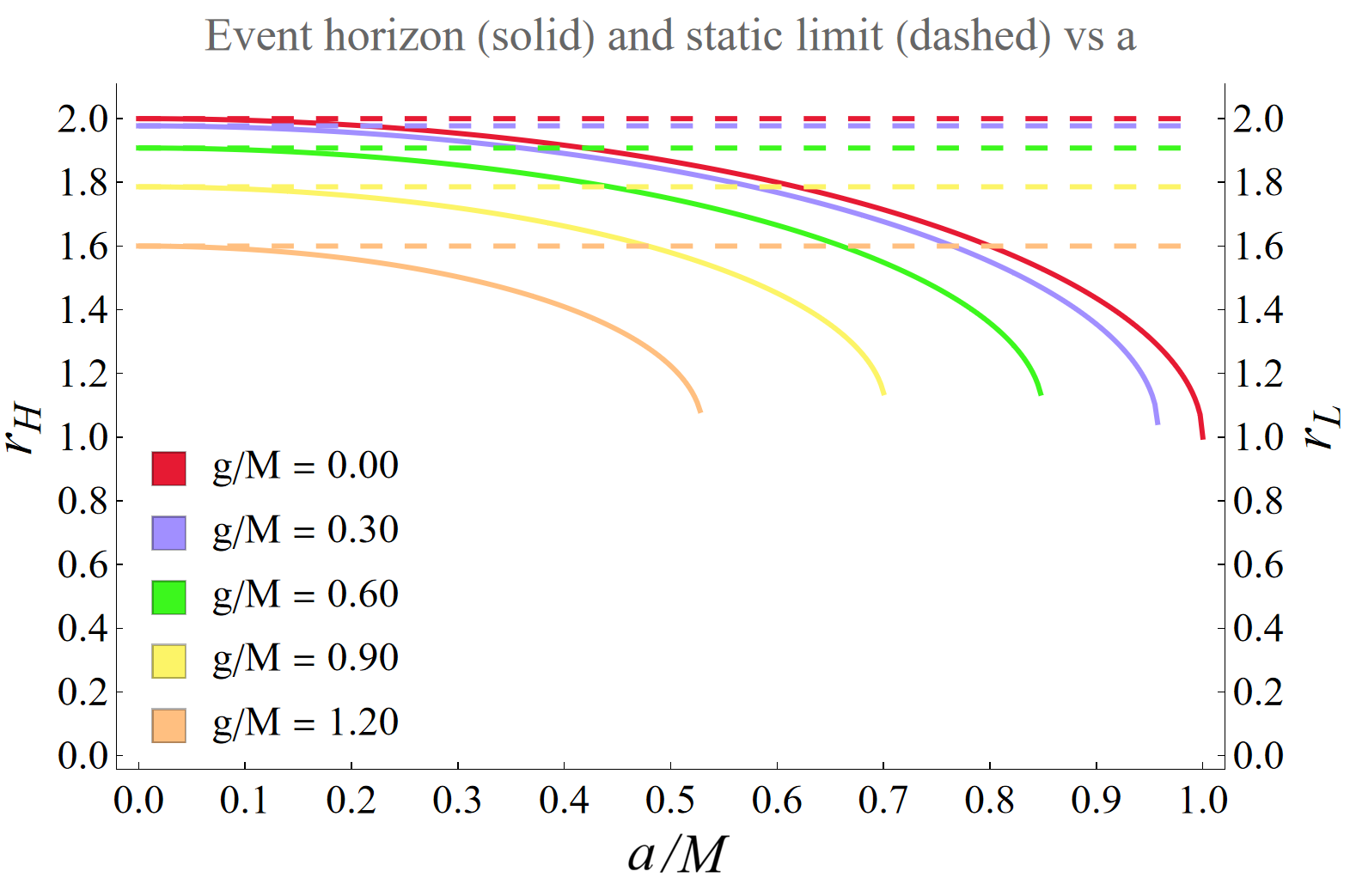}
  \caption{\textbf{Outer horizon and equatorial static limit versus spin.}
  Outer-horizon radius \(r_H\) (solid) and equatorial static-limit radius \(r_L\) (dot--dashed) as functions of the spin \(a/M\), for representative values of \(g/M\) within the black-hole domain.}
  \label{fig:rherL-vs-a}
\end{figure}

\begin{figure}[!htbp]
  \centering
  \includegraphics[width=\columnwidth,height=5.5cm,keepaspectratio]{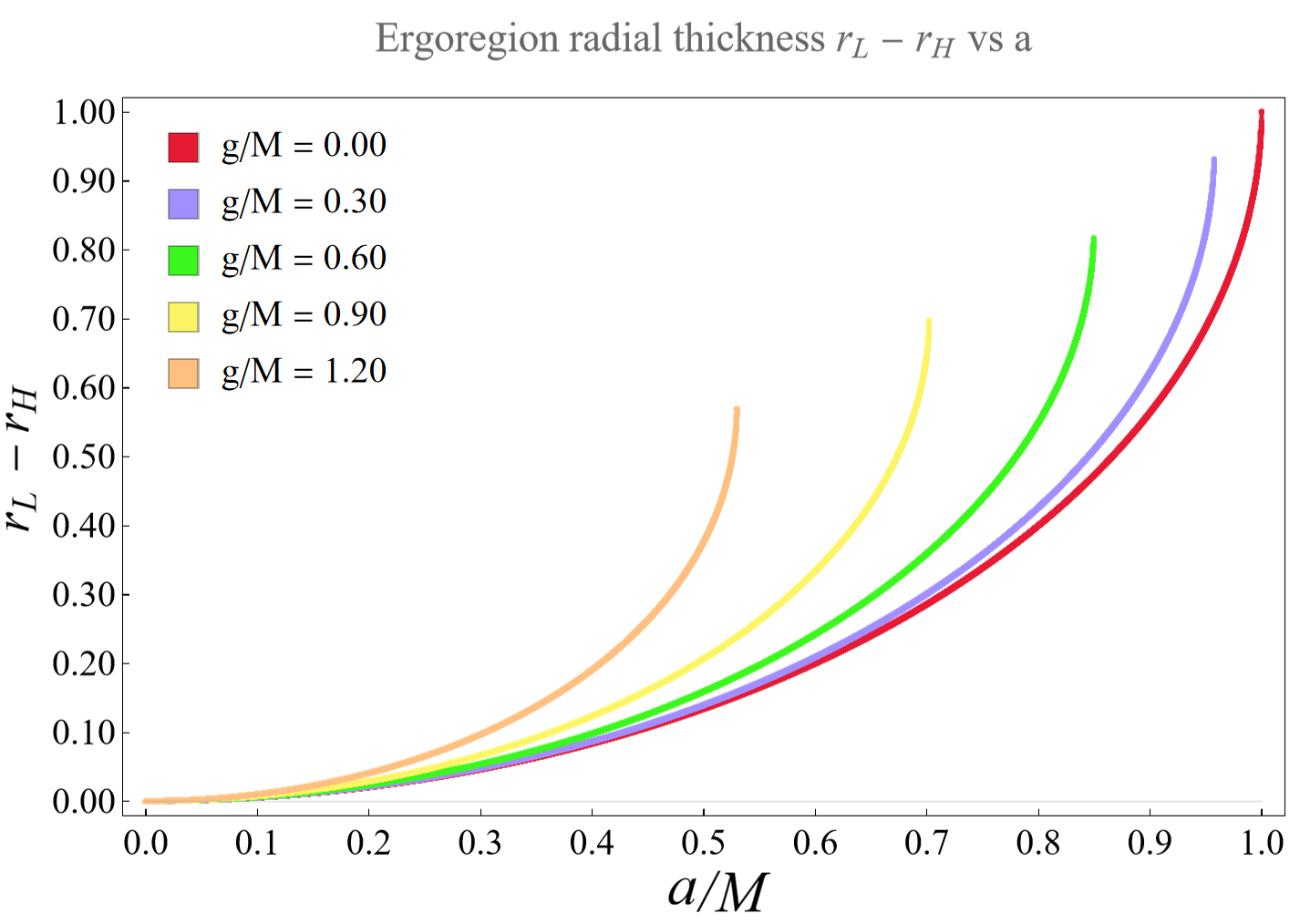}
  \caption{\textbf{Equatorial ergoregion thickness.}
  Radial thickness \(\Delta r(a;g)=r_L-r_H\) of the equatorial ergoregion as a function of \(a/M\), for the same representative values of \(g/M\) used in Fig.~\ref{fig:rherL-vs-a}. Each series is shown only where the outer horizon exists.}
  \label{fig:ErgoThickness}
\end{figure}

The outer horizon \(r_H\) defines the inner edge of the ergoregion, while the equatorial static limit \(r_L\) defines its outer boundary. Their behavior as functions of the spin \(a/M\), for representative values of the deformation parameter \(g/M\), is shown in Fig.~\ref{fig:rherL-vs-a}. The associated equatorial ergoregion thickness,
\begin{equation}
\Delta r(a;g)\equiv r_L(a;g)-r_H(a;g),
\end{equation}
is displayed in Fig.~\ref{fig:ErgoThickness}. Each curve is plotted only in the portion of parameter space where the outer horizon exists, so that \(\Delta r\ge 0\) remains well defined and physically meaningful.

\medskip

\begin{figure*}[t]
    \centering

    \begin{minipage}[t]{0.49\textwidth}
        \centering
        \includegraphics[width=\linewidth]{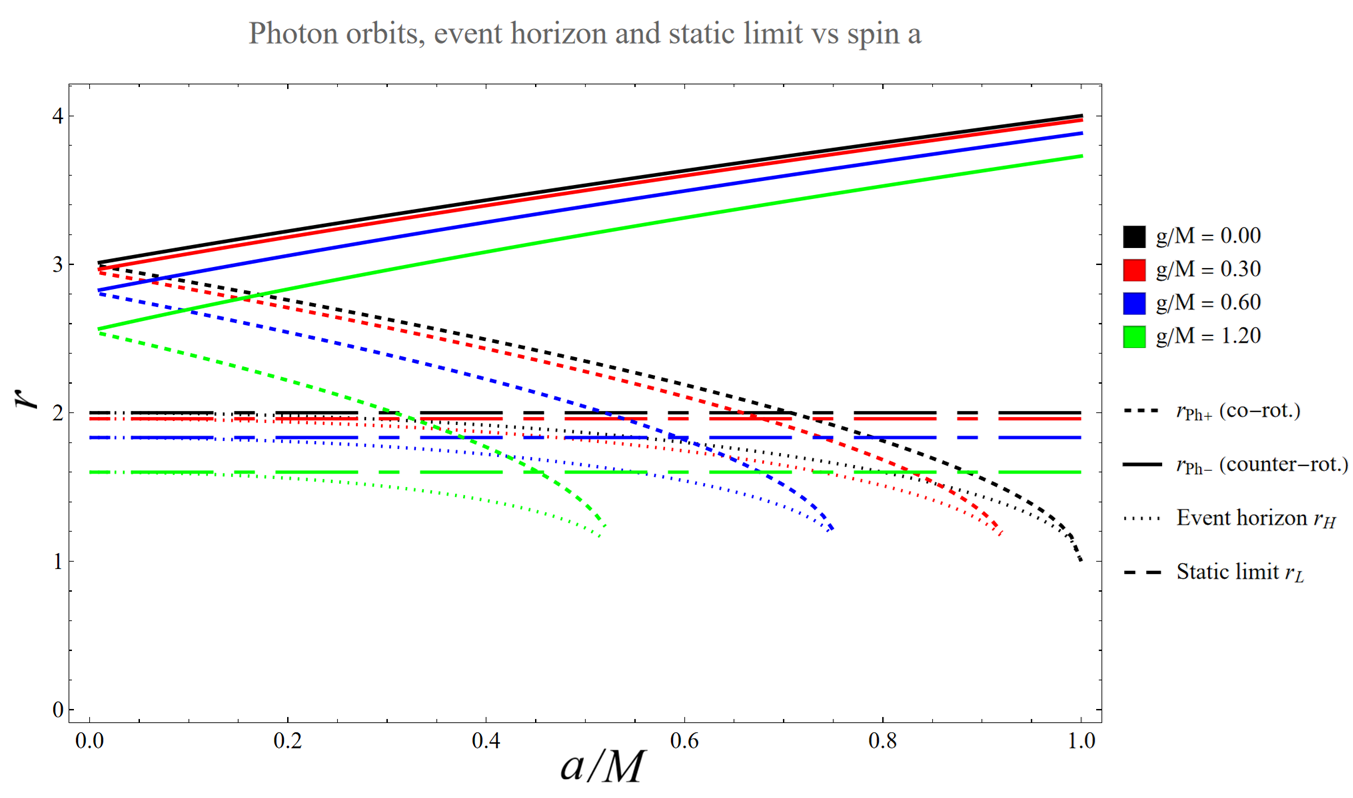}
        \caption{\textbf{Equatorial circular photon orbits.}
        Prograde (dashed) and retrograde (solid) circular photon
        radii as functions of the spin \(a/M\), for representative
        values of \(g/M\). Dotted guides indicate the equatorial
        outer horizon \(r_H\) and static-limit radius \(r_L\).}
        \label{fig:photonshpherada}
    \end{minipage}
    \hfill
    \begin{minipage}[t]{0.49\textwidth}
        \centering
        \includegraphics[width=\linewidth]{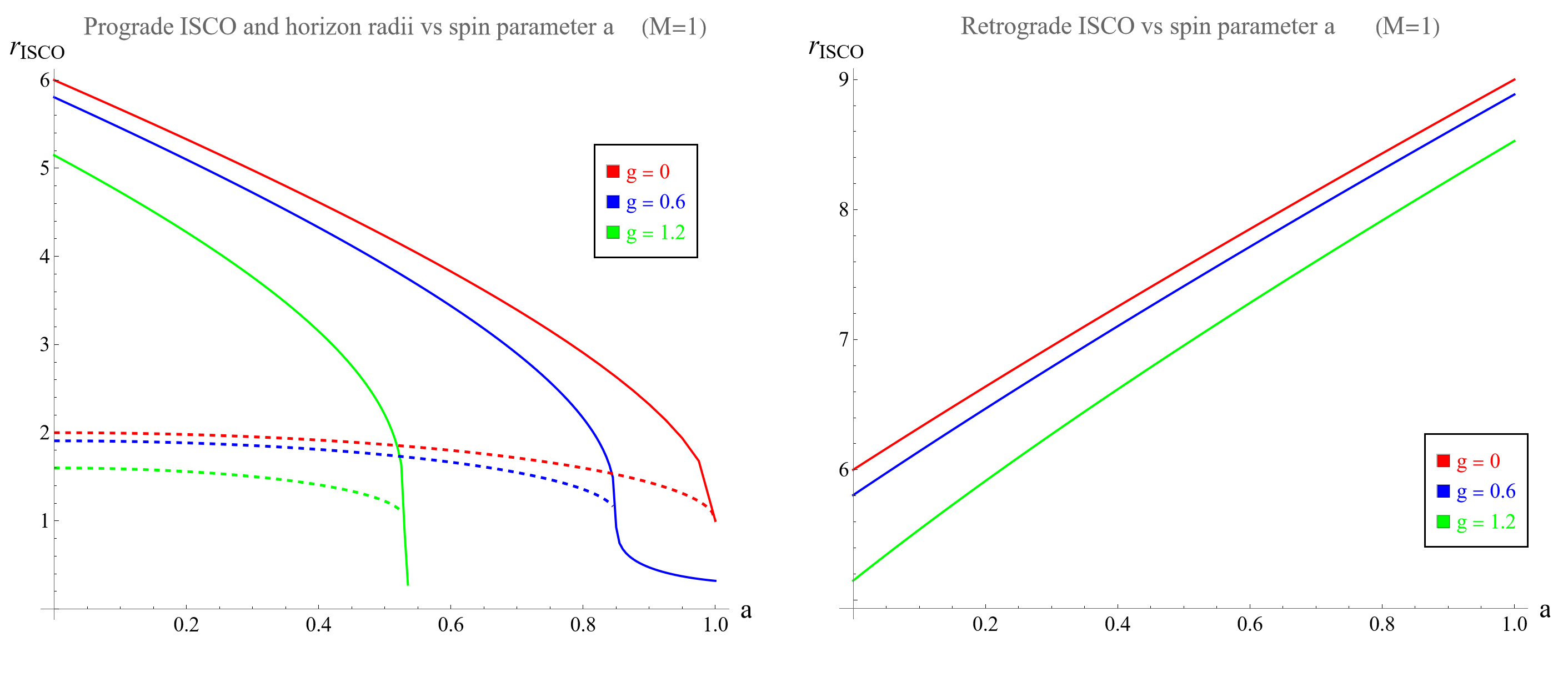}
        \caption{\textbf{ISCO radius versus spin.}
        ISCO radius \(r_{\mathrm{ISCO}}\) as a function of the
        spin \(a/M\), for representative values of the NED
        parameter \(g/M\). The outer-horizon radius \(r_H\)
        is overplotted as a dashed guide where appropriate,
        and each curve is truncated when the horizon
        ceases to exist.}
        \label{fig:riscoa}
    \end{minipage}

\end{figure*}

Additional insight is provided by equatorial geodesic diagnostics~\cite{DSSG,Wald}. Circular null orbits are obtained from the standard circularity conditions \(V_{\rm eff}=V_{\rm eff}'=0\) with \(k=0\), while the ISCO of timelike equatorial motion is determined by
\begin{equation}
V_{\rm eff}=0,\qquad V_{\rm eff}'=0,\qquad V_{\rm eff}''=0.
\end{equation}
The corresponding equatorial circular photon radii are shown in Fig.~\ref{fig:photonshpherada}. As expected, frame dragging produces the familiar prograde/retrograde asymmetry, while the NED deformation shifts both branches with respect to the Kerr limit. The behavior of the ISCO radius is summarized in Fig.~\ref{fig:riscoa}, where the curves are truncated whenever the event horizon ceases to exist.

\section{Explicit expressions of the $\{{\dot t}, {\dot \phi}, {\dot r}\}$}
\label{Appendixdottr}

We report here the explicit expression of $\dot{t}$ and $\dot{\phi}$  in terms of the conserved quantities $E$ and $l_z$, which are the energy and the $z$-component of angular momentum per unit mass 

\begin{widetext}

\begin{align}
\dot{t} &= \frac{g_{\phi \phi} E + g_{t \phi} l_z}{g_{t \phi}^2 - g_{tt} g_{\phi \phi}} = \frac{A \sin^2 \theta E - \frac{2 a M r^2}{\sqrt{r^2+g^2}}l_z }{\frac{4a^2 M^2 r^4 }{\Sigma (r^2 +g^2)} + A \sin^2 \theta \left( 1 - \frac{2M r^2 }{\Sigma \sqrt{r^2 + g^2}}\right)  }=\\ \notag
&\textcolor{black}{-\frac{(r^2 + a^2 \cos^2 \theta) \left(-\frac{2 a l_z M r^2}{\sqrt{g^2 + r^2}} + 
    E \left((a^2 + r^2)^2 - 
      a^2 \left(a^2 + r^2 \left(1 - \frac{2 M}{\sqrt{g^2 + r^2}}\right)\right) \sin^2 \theta\right)\right)}{\frac{4 a^2 M^2 r^4 \sin^2 \theta}{g^2 + r^2} - \left(r^2 \left(-1 + \frac{2 M}{\sqrt{r^2 + g^2}}\right) - a^2 \cos^2 \theta\right) \left((a^2 + r^2)^2 - 
      a^2 \left(a^2 + r^2 \left(1 - \frac{2 M}{\sqrt{g^2 + r^2}}\right)\right) \sin^2 \theta\right)}} \\ \vspace{12 pt}
\dot{\phi} &= -\frac{g_{t \phi} E + g_{tt} l_z}{g_{t \phi}^2 - g_{tt} g_{\phi \phi}} = \frac{\frac{2 a M r^2}{\sqrt{r^2+g^2}}E + \left(\frac{2M r^2 }{\Sigma \sqrt{r^2 + g^2}}-1\right)l_z }{\frac{4a^2 M^2 r^4 }{\Sigma (r^2 +g^2)} + A \sin^2 \theta \left( 1 - \frac{2M r^2 }{\Sigma \sqrt{r^2 + g^2}}\right)  }= \\ \notag 
&\textcolor{black}{\frac{(r^2 + a^2 \cos^2 \theta) \csc^2 \theta \left(-a^2 l_z + l_z r^2 \left(-1 + \frac{2 M}{\sqrt{r^2 + g^2}}\right) + a \left(a l_z + \frac{2 E M r^2}{\sqrt{g^2 + r^2}}\right) \sin^2 \theta\right)}{\frac{4 a^2 M^2 r^4 \sin^2 \theta}{g^2 + r^2} - \left(r^2 \left(-1 + \frac{2 M}{\sqrt{r^2 + g^2}}\right) - a^2 \cos^2 \theta\right) \left((a^2 + r^2)^2 - a^2 \left(a^2 + r^2 \left(1 - \frac{2 M}{\sqrt{g^2 + r^2}}\right)\right) \sin^2 \theta\right)}}
\end{align}

The radial equation of motion for null equatorial ($\theta=\pi/2\;\mathrm{and} \; \dot{\theta}=0$) geodesics in the rotating NED metric is: 
\begin{flalign}
    \dot{r} &=\pm \sqrt{-\frac{1}{g_{rr}} \left( g_{tt} \dot{t}^2 + 2 g_{t\phi}\dot{t}\dot{\phi}+g_{\phi\phi}\dot{\phi}^2 \right)}=\qquad
    &\qquad 
\end{flalign}
\textcolor{black}{
\tiny{
\begin{align*}
=& \pm \Bigg[ -\frac{\left(-2 M r^2 + (a^2 + r^2) \sqrt{g^2 + r^2}\right)}{\left(r^5 (g^2 + r^2 - 2 M \sqrt{g^2 + r^2}) 
+ a^2 (-4 M^2 r (-1 + r^2) + r^3 (g^2 + r^2))\right)^2} 
 \Bigg(4 a^3 E l_z M (g^2 r^2 + r^4 - 4 M^2 (-1 + r^2)) + 4 a E l_z M r^4 (g^2 + r^2 - 2 M \sqrt{g^2 + r^2}) \\ \notag
& \quad + r^8 (g^2 + r^2) \left(L_z^2 r^4 \sqrt{g^2 + r^2} - E^2 (-2 M + \sqrt{g^2 + r^2})\right) 
+ a^6 l_z^2 r^6 \left(8 M^3 + 6 M r^2 + 12 M^2 \sqrt{g^2 + r^2} + r^2 \sqrt{g^2 + r^2} + g^2 (6 M + \sqrt{g^2 + r^2})\right) \\ \notag
& \quad + a^2 \Bigg(-2 E^2 r^4 \sqrt{g^2 + r^2} (g^2 r^2 + r^4 + M^2 (2 - 4 r^2)) 
+ l_z^2 \left(8 M^3 + 6 M r^{10} (g^2 + r^2) + 3 r^{10} (g^2 + r^2)^{3/2} - 4 M^2 \sqrt{g^2 + r^2} (1 + 2 r^4)\right)\Bigg) \\ \notag
& \quad + a^4 r^2 \Big(-E^2 (g^2 r^2 + r^4 - 4 M^2 (-1 + r^2)) (2 M + \sqrt{g^2 + r^2}) 
+ l_z^2 \left(-16 M^3 + 12 M r^6 (g^2 + r^2) + 3 r^6 (g^2 + r^2)^{3/2} + 4 M^2 \sqrt{g^2 + r^2} (-2 + 3 r^6)\right) \Big) \Bigg) \Bigg]^{1/2}
\end{align*}}}

\end{widetext}

\bibliographystyle{unsrt}
\bibliography{biblio}

\end{document}